\documentclass[journal]{IEEEtran}

\ifCLASSINFOpdf

\usepackage{cleveref}
\usepackage{dirtytalk}
\usepackage[colorinlistoftodos,prependcaption,textsize=medium]{todonotes}
\usepackage{color}
\usepackage[printonlyused,nohyperlinks]{acronym}
\usepackage{subfig}
\usepackage{url} 
\usepackage{graphicx}
\usepackage{adjustbox}
\usepackage{caption}
\usepackage{enumitem}
\usepackage{blindtext}
\usepackage{comment}
\usepackage[subtle]{savetrees}

\usepackage{multirow}

\newcommand{\ie}[0]{\textit{i.e.}, } 
\newcommand{\eg}[0]{\textit{e.g.}, }

\newcommand{\etc}[0]{\textit{etc.}}
\newcommand{\vs}[0]{\textit{vs.}}

\renewcommand{\theenumi}{\Alph{enumi}}

\begin{document}

\title{Community Networks and Sustainability: a Survey of Perceptions, Practices, and Proposed Solutions}

\author{\IEEEauthorblockN{Panagiota Micholia\IEEEauthorrefmark{1}\thanks{}, Merkouris Karaliopoulos\IEEEauthorrefmark{1}, Iordanis Koutsopoulos \IEEEauthorrefmark{1}, Leandro Navarro \IEEEauthorrefmark{2},\\ Roger Baig \IEEEauthorrefmark{2}, Dimitris Boucas\IEEEauthorrefmark{3}, Maria Michalis \IEEEauthorrefmark{3}, Panayiotis Antoniadis\IEEEauthorrefmark{4} }\\
 
 \IEEEauthorrefmark{1} Athens University of Economics and Business ,
\IEEEauthorrefmark{2}Universitat Polit\`ecnica de Catalunya ,\\
\IEEEauthorrefmark{3}University of Westminster, \IEEEauthorrefmark{4}NetHood Zurich \\

\{panamixo,jordan, mkaralio\}@aueb.gr, \{leandro, rbaig\}@ac.upc.edu, \\ \{D.Boucas, M.Michalis\}@westminster.ac.uk, panayotis@nethood.org   

}


\maketitle


\IEEEpeerreviewmaketitle

\begin{abstract}
Community network (CN) initiatives have been around for roughly two decades, evangelizing a distinctly different paradigm for building, maintaining, and sharing network infrastructure but also defending the basic human right to Internet access. Over this time they have evolved into a mosaic of systems that vary widely with respect to their network technologies, their offered services, their organizational structure, and the way they position themselves in the overall telecommunications' ecosystem. Common to all these highly differentiated initiatives is the \emph{sustainability} challenge.
%
We approach sustainability as a broad term with an economical, political, and cultural context. 
We first review the different perceptions of the term. These vary both across and within the different types of stakeholders involved in CNs and are reflected in their motivation to join such initiatives. 
%
Then, we study the diverse ways that CN operators pursue the sustainability goal. Depending on the actual context of the term, these range all the way from mechanisms to fund their activities and synergistic approaches with commercial service providers, to organizational structures and social activities that serve as \emph{incentives} to maximize the engagement of their members.
%
Finally, we iterate and discuss theoretical concepts of \emph{incentive mechanisms} that have been proposed in the literature for these networks as well as implemented tools and processes designed to set the ground for CN participation. 
While, theoretical mechanisms leverage game theory, reputation frameworks, and social mechanisms, implemented mechanisms focus on organizational matters, education and services, all aiming to motivate the active and sustained participation of users and other actors in the CN. 
\end{abstract}
\section{Introduction}
\subsection{CNs: history, evolution and current status}

Community networks (CNs) are networks inspired, built and managed by citizens and non-profit organizations. They are crowdsourced initiatives where people combine their efforts and resources in a collective manner to instantiate communication network infrastructures. 

While the phenomenon of community initiatives in the field of media is as old as the distinct media themselves, CNs originally surfaced in the late 90s and have taken many forms and shapes ever since. Typically, these CNs are initiated by tiny groups of people, usually in the range of one to ten, who more often than not are driven by strong cultural and political motives. They contribute to the fight against the digital divide through the provision of telecommunication services in under-served areas, the desire for autonomy and self-organization practices, the right to open, neutral networks and privacy, the experimentation with technology in do-it-yourself manner, and the commitment to community ideals and needs. 

Whereas some CNs have become obsolete due to the rise of commercial high speed broadband networks in the areas CNs operated, others have flourished and evolved into alternative telecommunication network models (section \ref{subsec:additional}). Not only have they filled in the coverage gaps of commercial operators providing telecom services in rural areas, but they have also developed rich organizational frameworks with various tools and mechanisms. Typically, these frameworks emerge as a result of past experiences, successful and unsuccessful practices and accumulated knowledge. They are meant to systematize the network's governance, management and operation and ensure the CNs sustainability. The establishment of functional economic models is a key factor to this end. 

\subsection{Current motivating factors and new paths for CNs}\label{subsec:additional}
With a few notable exceptions (e.g., \cite{Baig2015150}), most community networks have been viewed (and have been viewing themselves) as alternative networks that are incompatible with any commercial notion, not least because of the strong cultural/political values of the small groups that initiated them. 
Yet, there seem to currently exist additional good reasons that motivate a reiteration of their positioning in the overall telecommunications landscape and new approaches to their sustainability.

\begin{figure*}
\vspace{-1cm}
     \centering
     \subfloat[][]{\includegraphics[width=7.3cm]{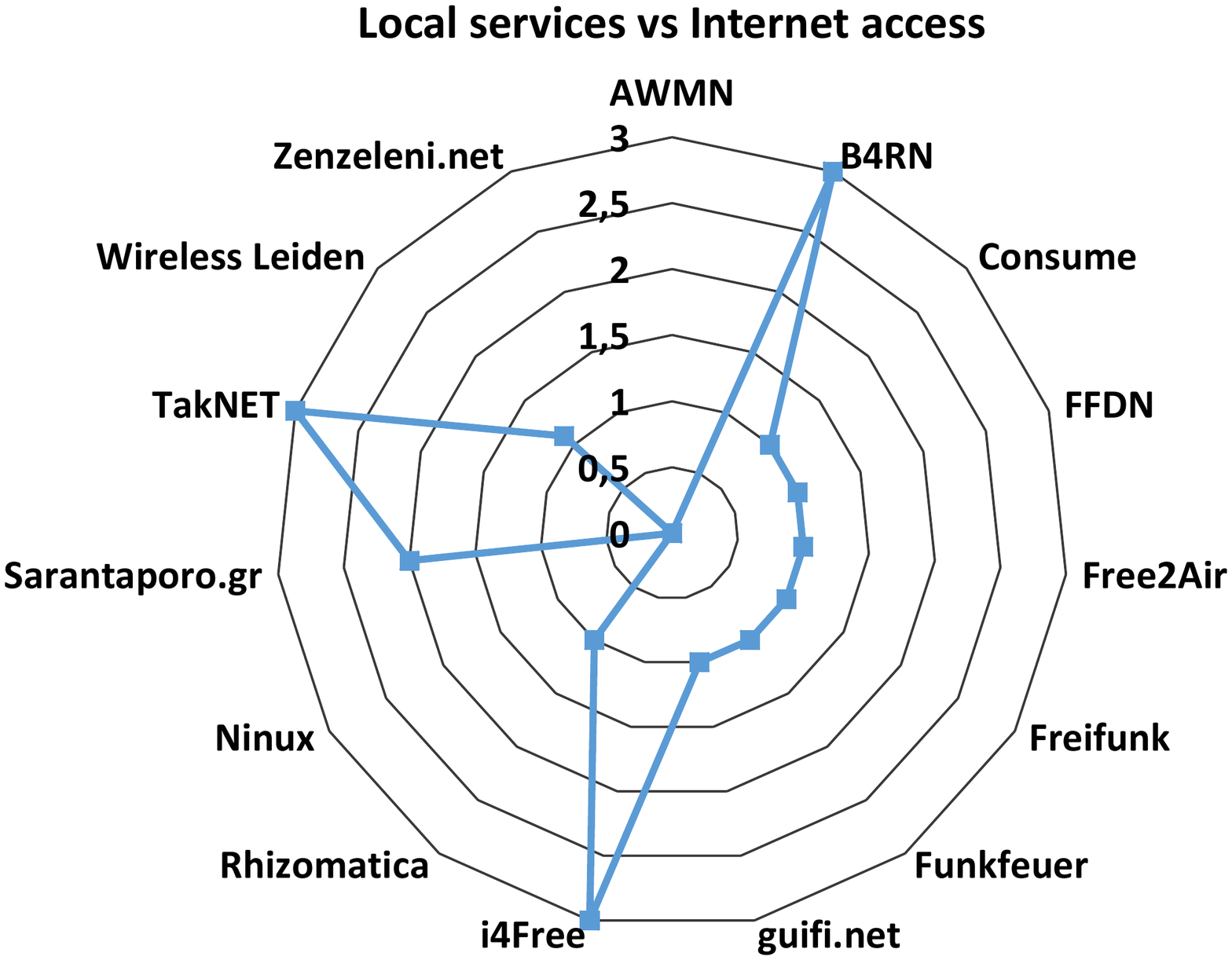}\label{<figure1>}}
     \hspace{-1.99cm}
     \subfloat[][]{\includegraphics[width=7.3cm]{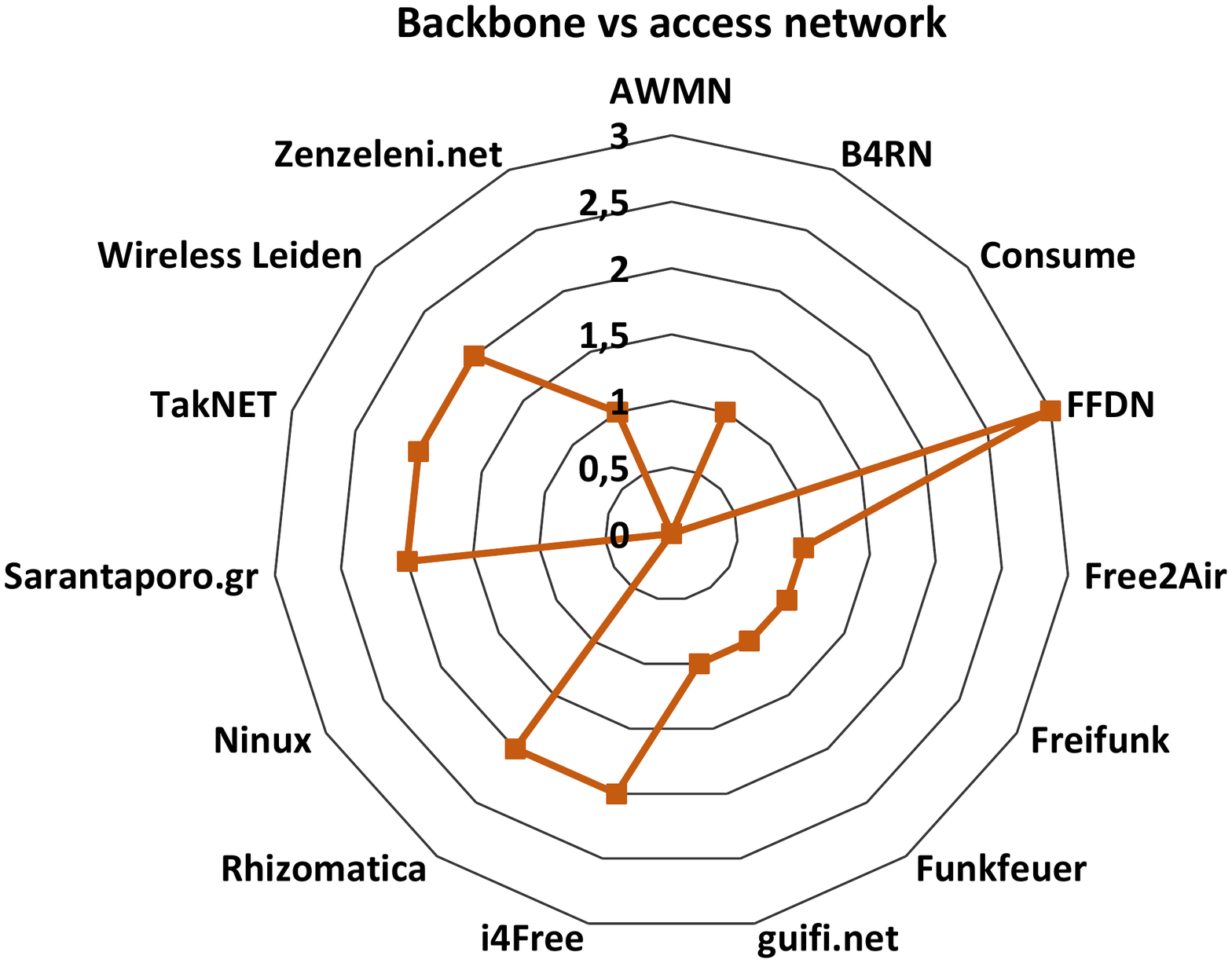}\label{<figure2>}}
     \hspace{-1.99cm}
     \subfloat[][]{\includegraphics[width=7.3cm]{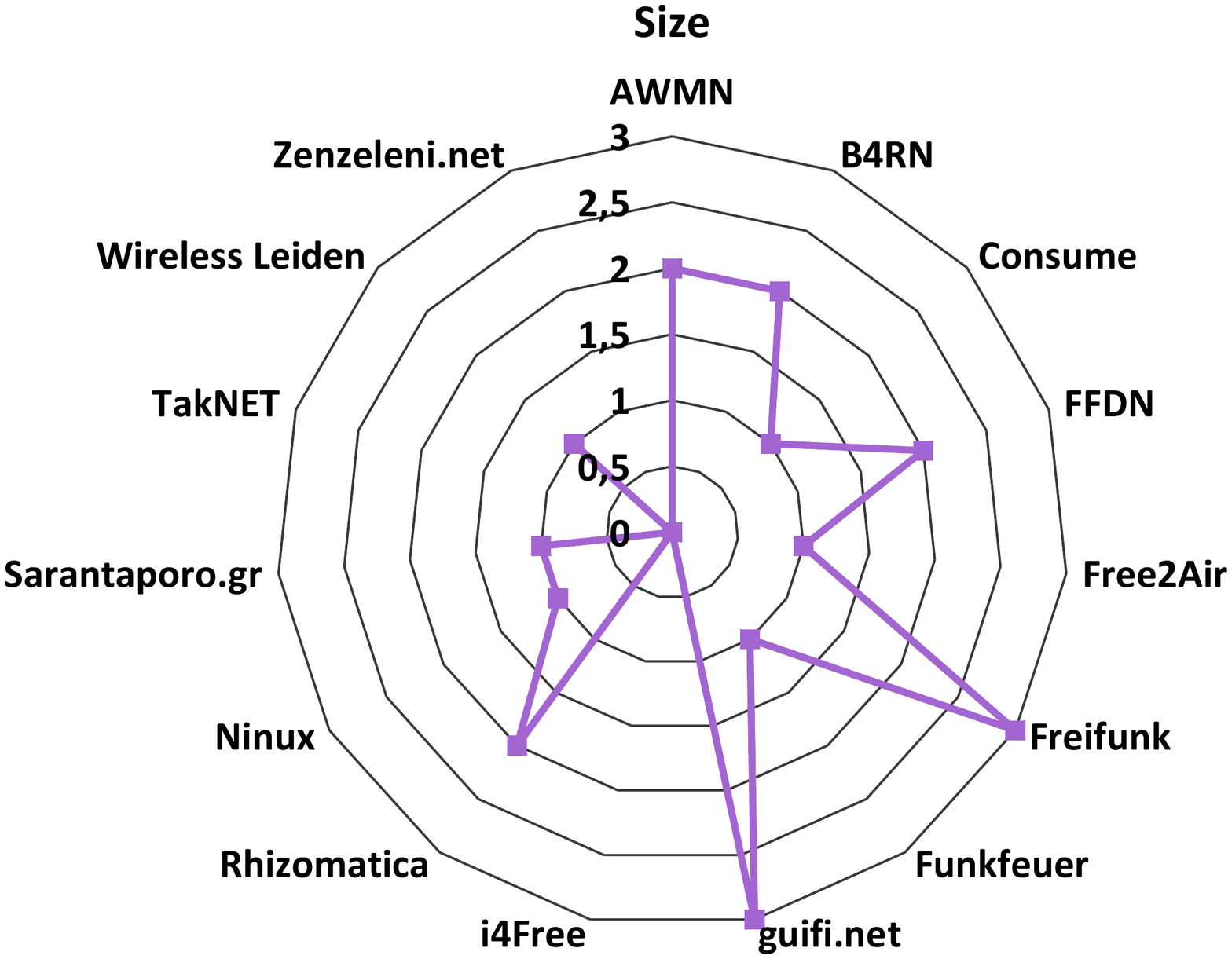}\label{<figure3>}}
    \caption{Radar charts following CN characteristics \ie types of services and infrastructure used, number of participating nodes. (a) \textit{CN services}. 0: local services as default (Internet connectivity available upon request, manual configuration), 1: mix of local services and Internet connectivity, 2: Internet connectivity as the main service (only management tools as local services), 3: Internet connectivity only. (b) \textit{CN Infrastructure}. 0: mostly backbone network, access points offered by certain individuals through their home routers, 1: mostly backbone, good access points, 2: mostly access network (backbone is used to connect to the Internet or interconnect small "islands"), 3: only access network (no backbone network).  
    (c) \textit{CN Size}. 0: Very small (number of nodes$ <100$), 1: Small ($100 <$ number of nodes $< 1000$), 2: Medium ($1000<$ number of nodes $<10.000$), 3: Large (number of nodes $>10.000$).}
     \label{fig:chart}
\end{figure*}

\vspace{2mm}
\subsubsection{Contributing to broadband connectivity goals}
Broadband Internet access has been promoted as a core priority of political agendas throughout the world. In Europe, for example, the European Commission (EC) has set ambitious policy objectives for the years to come, summarized under the EC broadband 2020\footnote{\url{https://ec.europa.eu/digital-single-market/en/broadband-strategy-policy}} and 2025\footnote{\url{https://ec.europa.eu/digital-single-market/en/broadband-europe}} agendas. 
These agendas demand huge investment costs and grassroots initiatives such as CNs are acknowledged as one possible response to this challenge and one of four ways to involve public authorities in the realization of the broadband vision \cite{ecbb2014}. Community broadband networks such as the Catalan guifi CN \cite{Baig2015150} are singled out in the EC website as best practices in this respect.

\vspace{2mm}
\subsubsection{Realizing Internet access in developing regions}
More than half of the world's population --specifically women, the poor and marginalised populations in developing areas-- are still offline~\cite{ITUBR16}. Many large industrial corporations such as Google, Microsoft and Facebook have stated ambitious objectives for connecting another billion users around the globe. These initiatives are commercial, for-profit and often do not plan access to the open Internet. Combining the Do-It-Yourself culture with provisions for unlicensed spectrum and cheap fibre, small crowdfunded community operators that create local value for the local people, without need for complex and centralized systems, may be the obvious way to go about realizing the vision of Internet access to developing regions. 

\begin{center}
\begin{table*}
\centering
\resizebox{18.4cm}{!} {
    \begin{tabular}{ | p{1.8cm} |  p{1.8cm} |  p{2cm} | p{1.2cm} |p{9cm} |}
    \hline
   \multirow{2}{*}{\textbf{CN}} &\multirow{2}{*}{\textbf{Location}} & \textbf{Networking technology} & \multirow{2}{*}{\textbf{Internet}} & \multirow{2}{*}{\textbf{Description}}
   
   \\ \hline
   
\multirow{3}{*}{AWMN}  & \multirow{3}{*}{Greece}   & \multirow{3}{*}{wifi}  & \multirow{3}{*}{Yes*} & Built by network technicians, enthusiasts and radio amateurs. Contains native services without need for public Internet connectivity \ie
    games, libraries, network monitoring tools, DNS solutions,  and experimental platforms. (2002)
\\ \hline

\multirow{2}{*}{B4RN} & \multirow{2}{*}{UK} & \multirow{2}{*}{fibre} & \multirow{2}{*}{Yes} & Started by a local volunteer, who led the group as a networking expert. Aimed at bridging the digital
divide. Based exclusively on fiber.(2011)
\\ \hline

   \multirow{3}{*}{Consume} & \multirow{3}{*}{UK} & \multirow{3}{*}{wifi} & \multirow{3}{*}{Yes} & One of the first CNs to be conceived
and deployed in Europe. The original motivation was to save Internet access 
fees for conducting business. It has epitomised the anti-commercial model of networking. Not active anymore. (2000)
\\ \hline

\multirow{2}{*}{FFDN} &  \multirow{2}{*}{France, Belgium} & \multirow{2}{*}{wifi, DSL/fibre} & \multirow{2}{*}{Yes} & An umbrella organization
embracing 28 CNs operating across France. Adheres to values
of collaboration, openness and support of human rights
(freedom of expression, privacy). (2011)
\\ \hline 

\multirow{2}{*}{Free2Air} & \multirow{2}{*}{ UK} & \multirow{2}{*}{wired, wifi} & \multirow{2}{*}{Yes} & An alternative to the commercial Internet
provision. Run by a small number
of artists and a number of other individuals until 2015. (1999)
\\
\hline
  
\multirow{2}{*}{Freifunk} &  \multirow{2}{*}{Germany} & \multirow{2}{*}{fibre, wifi}  & \multirow{2}{*}{Yes} & An open initiative that supports free computer
networks in Germany. It attracted many
artists, activists and tech enthusiasts from all over
Europe. (2002)
\\ \hline

\multirow{2}{*}{Funkfeuer} &  \multirow{2}{*}{Austria} & \multirow{2}{*}{wireless}  &  \multirow{2}{*}{Yes} & A free experimental wireless
network across Austria, committed
to the idea of DIY, built and currently
maintained by a group of computer enthusiasts. (2003)
\\ \hline  

\multirow{2}{*}{guifi.net} & \multirow{2}{*}{Spain } &  \multirow{2}{*}{fibre, wifi} &  \multirow{2}{*}{Yes*} & Started in Osona to serve remote rural areas that were not covered by conventional ISPs. Applies the principles of CPR management. (2004)
\\ \hline 

\multirow{2}{*}{i4Free} &  \multirow{2}{*}{Greece} & \multirow{2}{*}{wifi} & \multirow{2}{*}{Yes}  & The initiative of a German engineer and professor in an island of Greece with poor Internet connectivity. (2014)
\\ \hline

\multirow{2}{*}{Ninux} &  \multirow{2}{*}{Italy}  &  \multirow{2}{*}{wifi} & \multirow{2}{*}{No} & Experimentation and hacking culture. Ninux operates as an experimental
platform for decentralized protocols, policies and technologies. (2003)
\\ \hline

\multirow{3}{*}{Rhizomatica} & \multirow{3}{*}{Mexico}  & \multirow{3}{*}{wireless} &\multirow{3}{*}{Yes} & Provides GSM services. Creates open-source technology and helps communities to build their own networks. Initiated by a small group of people with knowledge of community organization and technology.  (2009)
\\ \hline 

\multirow{2}{*}{Sarantaporo.gr} & \multirow{2}{*}{Greece}  & \multirow{2}{*}{wireless} &\multirow{2}{*}{Yes} & People with origins from the area
of Sarantaporo wanted to create a
website for their village when they
realized that there was no network connection. (2010)
\\ \hline 

\multirow{3}{*}{TakNET} & \multirow{3}{*}{Thailand} & \multirow{3}{*}{wifi} & \multirow{3}{*}{Yes} & Established as an academic project at the Asian Institute of Technology (AIT). Follows the goal of bridging the digital divide in Thailand villages. Composed of TakNET1, TakNET2 and TakNET3. (2012) 
\\ \hline

\multirow{3}{*}{Wireless Leiden} & \multirow{3}{*}{Netherlands} & \multirow{3}{*}{wifi} & \multirow{3}{*}{Yes} & Volunteer-based open, inexpensive, fast wireless network in Leiden and surrounding villages. Developed by a group of local residents. Provides Internet access and free local communication. (2002)
\\ \hline

\multirow{3}{*}{Zenzeleni.net} & \multirow{3}{*}{South Africa} & \multirow{3}{*}{wifi} & Yes, VoIP public phones & Initiated by researchers from the University of the Western Cape (UWC) in the rural under-developed area of Mankosi. Solar powered network. Operated as an umbrella co-operative enterprise and a telecoms provider. (2013) 
\\ \hline

\end{tabular}
}
\caption{Basic information about the 15 CN instances that are analyzed further in the survey. These are chosen as representative instances of the rich variety of worldwide CNs.}
\end{table*}
\label{tab:CN_list}
\end{center}
\vspace{-0.7cm}

\subsubsection{Democratization of the telecommunication market}
The market of telecom services is usually composed of monopolies and oligopolies that concentrate significant amount of power. The prevention of telecommunications market distortions and the openness of networks is a key goal set by the International Telecommunication Union (ITU)~\cite{itu2008}, the EC~\cite{ecbb2014}, and the Organization for Economic Cooperation and Development (OECD)~\cite{OECDreport}. Monopolies lead to vertically integrated models, where all the layers of the network belong to one entity and end users are left with limited options when it comes to choosing an operator.

The way they are built and operated makes CNs an ideal candidate model for separating the network infrastructure from the service provision layer. This separation generates opportunities for sharing the related costs between multiple players and opening the network to public administrations and commercial entities such as local/regional ISPs (we elaborate on this model in section II.C).

\vspace{0.4cm}
Our survey does not aim at presenting the status of the hundreds of CN efforts around the globe, nor is it a review of the technologies used in CNs today. Such information is already available in the CN literature~\cite{szabo2007wireless},~\cite{lawrence2007wireless},~\cite{5762819}. Instead, the focus of this survey is on the multiple, often complementary, ways different CN initiatives pursue their sustainability. We approach sustainability as a multi-faceted term, with technical, economic, socio-cultural and political context. We review how these networks fund their activities; which ones have been the dominant motives behind their initiation and which ones are the aspirations of other actors when participating in them; and what kind of tools and processes are in place as incentives in the different CNs to best respond to these motives and aspirations.

Most of the material for this survey originates from interviews, both in-person and questionnaire-based, carried out in the context of the netCommons R\&D project~\cite{NETCOMMONS_D2_2},~\cite{NETCOMMONS_D2_1}. Another big part, on proposed participation incentives and mechanisms, is the result of an exhaustive review of the existing scientific literature on the topic. Fifteen CNs are primarily discussed in this paper, as listed in Table \ref{tab:CN_list}. 
They are selected as good representatives of the diversity in existing CNs with respect to size, supported services (local services~\vs~Internet access), network scope/role (backbone network \vs~access network), geographical area of coverage (urban areas with rich communication alternatives~\vs rural under-served areas), organizational structure (involved actors and decision-making processes), and funding sources.  
The radar chart of Fig. 1 depicts how these fifteen CNs score on the first three attributes (size, services, network role) on a 0-3 scale.

In the remainder of the survey, we first present the layered network infrastructure model, which aims at maximal openness and involvement of actors, and explore how CNs fit in it as open access network instances (section \ref{sec:netinfra}). Then, in section \ref{sec:CN_sustain_incentives}, we iterate on the participation motives of different actors 
and their implications for the CN sustainability. In section \ref{sec:funding}, we elaborate on the economic sustainability aspects and the funding sources of CNs. Finally, we investigate actual and theoretical practices adopted by CNs or proposed in the literature (section \ref{mechanisms}), before concluding in section \ref{closure} with a list of the most valuable insights out of the survey.

\section{Network Infrastructures and \\ Community Networks}\label{sec:netinfra}

In order to understand how CNs fit in the broader picture of broadband networks, we contemplate the typical network infrastructure layers, the basic actors and the business models --as they are met in most networks-- below. 


{ \renewcommand{\arraystretch}{1.2}
\begin{center} 

\begin{table}[t]
\centering
\resizebox{8.5cm}{!} {
\begin{tabular}{|p{1.1cm}|p{6.7cm}|}
\hline
\textbf{Acronym} & \textbf{Description}  \\ \hline
AP & Access Point \\ \hline
CAPEX & Capital Expenditure \\ \hline
CN &  Community Network \\ \hline
CONFINE & Community Networks Testbed for the Future Internet \\ \hline
CPR &   Common Pool Resource  \\ \hline 
CS &  Community Service \\ \hline 
DIY & Do-It-Yourself \\ \hline  
DNS &  Domain Name Server \\ \hline
EC & European Commission \\ \hline
EU & European Union \\ \hline
GFOSS & Greek Free/Open Source Software Society \\ \hline 
ICT  & Information and Communication Technology \\ \hline
ISP & Internet Service Provider \\ \hline
ITU & International Telecommunications Union \\ \hline
MANET &  Mobile Ad-hoc Networks \\ \hline
NCL & Network Commons License \\ \hline
NP & Network Provider \\ \hline
NPO & Non-Profit Organization \\ \hline
OECD & Organization for Economic Cooperation and Development \\ \hline
OPEX & Operational Expenditure \\ \hline
P2P & Peer to Peer \\ \hline
P2PWNC &  Peer-to-peer Wireless Network Confederation \\ \hline
PIP & Physical Infrastructure Provider \\ \hline
SP & Service Provider \\ \hline
VoIP & Voice over Internet Protocol \\ \hline
WCED & World Commission Environment and Development \\ \hline
WCL & Wireless Commons License \\ \hline
\end{tabular}
}
\caption{Terminology used throughout the paper.}
\end{table}
\label{tab:acronyms}
\end{center}
}
\vspace{-1cm}
\subsection{Network Infrastructure Layers} 

Considering how a broadband network is created, its structure can be decomposed into three distinct but inter-dependent layers: a) \textit{passive infrastructure}, b) \textit{active infrastructure} and c) \textit{services}.

\begin{figure}[tbhp]
\centering
 \includegraphics[width=0.8\linewidth]{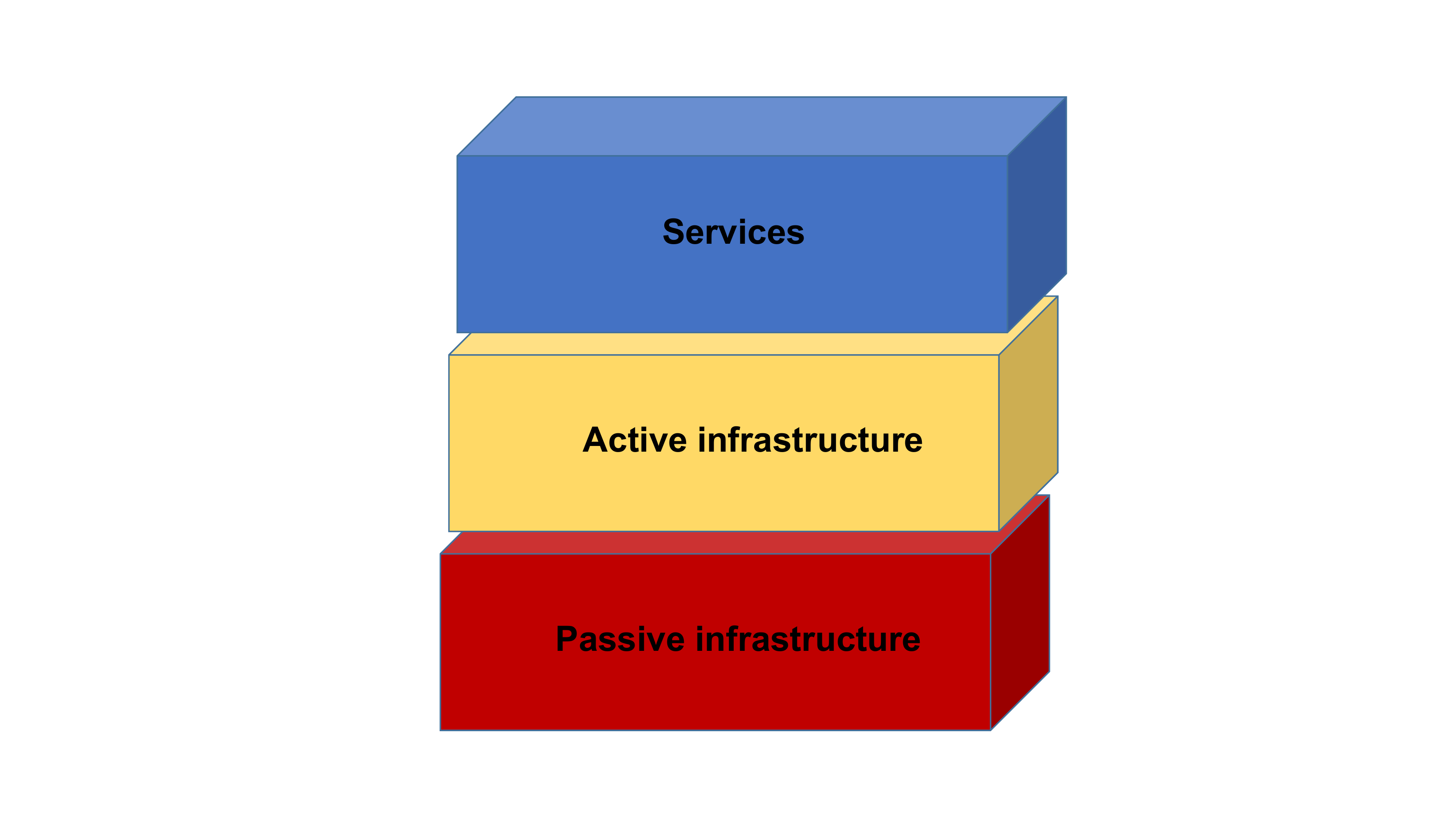}
 \caption{The layers of a broadband network.}
 \label{fig:broadbmodel_layers}
\end{figure}

The \textit{passive infrastructure layer} expresses the non-electronic physical equipment needed to deploy the network. Non-electric elements vary depending on the link technology in use, \eg fibre, copper, antennas. They typically refer to ducts, cables, masts, towers, technical premises, easements \etc   The passive infrastructure is built to endure for many years, usually decades. Its development demands high capital expenditure (CAPEX) and frequent upgrades are difficult to realize. However, its operational costs (OPEX) are relatively low. 

The second layer, \ie \textit{active infrastructure}, describes the electronic physical equipment of a network such as routers, switches, transponders, control and management servers. The OPEX of the active equipment is high (\ie electricity costs) but its capital expenditure is usually low since it involves up-to-date technological elements. The active equipment needs to follow the rapid advances of technology and get renewed frequently, \ie within a decade. 

The third and highest layer of a broadband network is the  layer of \textit{services}. It corresponds to the telecommunication services provided on top of the passive and active infrastructure. These services may be both private and public and include electronic government, education, health, commerce, Internet, entertainment, telephony (e.g., VoIP), access to media content (television, radio, movies) and many more. End users usually pay a fee for receiving the services either directly or indirectly. The type of reimbursement depends on the chosen network infrastructure model and the business actors involved. 

The implementation of the service layer is conditioned on the deployment of the passive and active infrastructure. Therefore, the first two layers are a prerequisite for the existence of the third one (Fig.\ref{fig:broadbmodel_layers}). 
 
\subsection{Business actors} 
Business actors are determined in accordance with the network infrastructure layers \cite{ecbb2014},\cite{NETCOMMONS_D1_2},\cite{Forzati2010},\cite{szabo2007wireless}. They are typically providers of the network's equipment and services. Telecom operators and private companies, public authorities, local cooperatives and housing associations, are some characteristic examples of business actors.  

In detail, the \textit{physical infrastructure provider (PIP)} has ownership of the passive equipment and undertakes the equipment's maintenance and operation responsibilities. PIPs can be divided into \textit{backbone PIPs} and \textit{access area PIPs}, depending on which network parts they possess. Backbone PIPs invest in the backbone network infrastructure, while access area PIPs own and moderate the infrastructure aimed for providing connections to the end users \ie first-mile connectivity. In the case of CNs, a local organization may participate as a backbone PIP, an access PIP or both. 

The \textit{network provider (NP)} owns and operates the active equipment. It leases physical infrastructure installations from the PIPs and makes its equipment available for the provision of services by other SPs or provides its own services. Network providers may be public authorities, private companies, local cooperatives who own the equipment or entities who are subcontracted to operate them by one of the aforementioned owner entities. 

The \textit{service provider} offers services within the network. Service providers are typically companies that utilize the network's active and passive equipment to offer their services to end users in exchange for compensation, typically payment. The payment can be direct (service fee) or indirect (connection or network fee). They need access to the NP's interface and install their own devices if and where needed. The existence of service provision within the network is vital for the end user engagement and therefore the network's viability. 

\subsection{Network Infrastructures Business Models}
\begin{figure}[tbhp]
\centering
\includegraphics[width=1\linewidth]{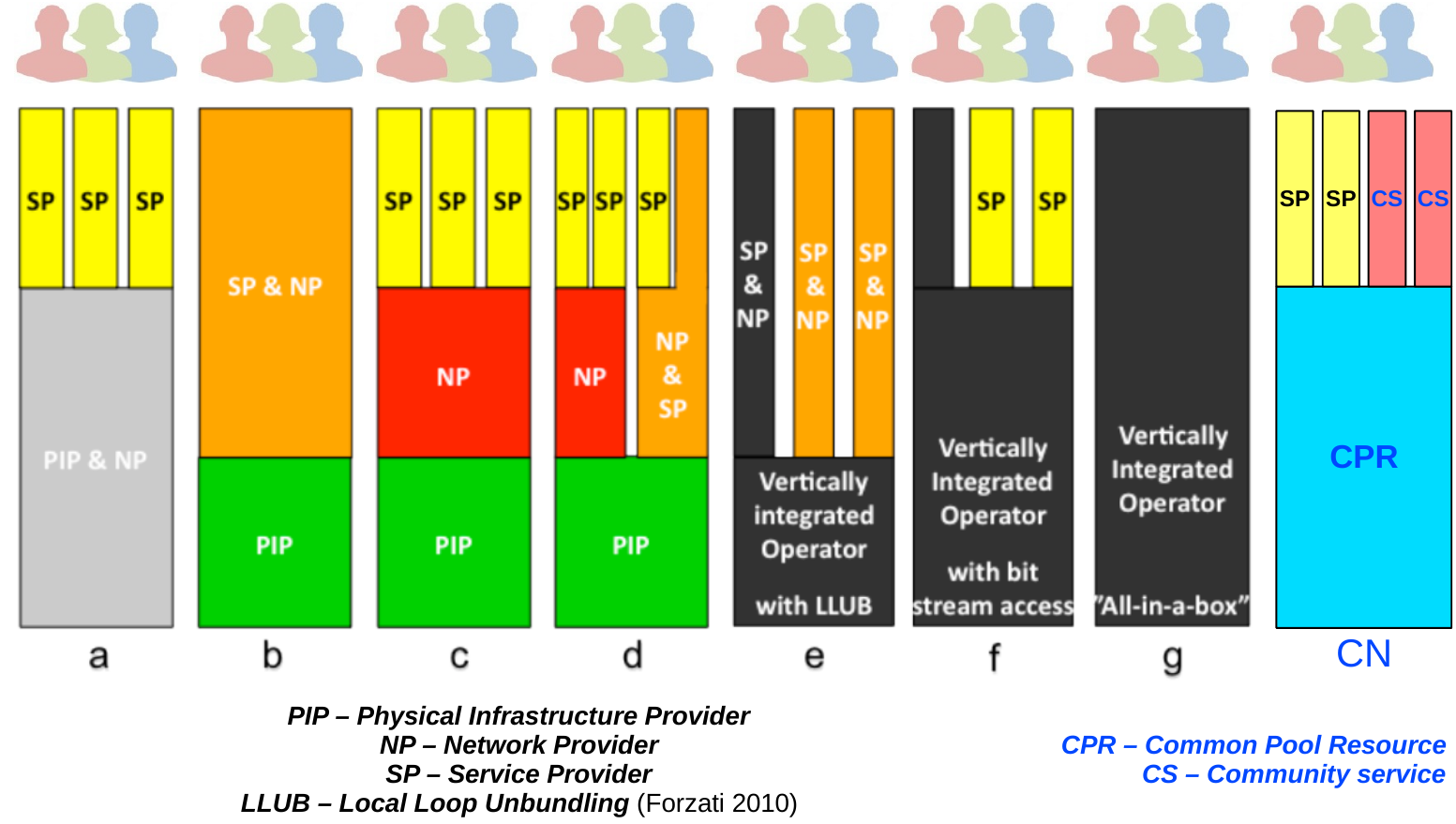}
\caption{The components of a broadband network (with a focus on optical fibre) and the three service layers.}
\label{fig:broadbmodel}
\end{figure}

The roles and responsibilities of different business actors in network infrastructures vary resulting in a great range of business models (Fig. \ref{fig:broadbmodel}).
Traditional telecom models follow the concept of \textit{vertical integration}. In these models, the ownership and operation of all three infrastructure layers is concentrated to one single entity. As a consequence, cases of monopolies or oligopolies that hamper the existence of competitors by exercising great control over the market, \ie "market failure" cases, are common. Moreover, due to lack of other competing entities, a single vertical integrated operator is often not willing to provide broadband access to remote areas featuring high network expansion costs, leaving several rural areas under-served. 

To reverse this picture, the ITU \cite{ecbb2014},\cite{itu2008} and the EC have set a goal to promote infrastructure separation and sharing through legislation, regulation and subsidies. Open access networks have been brought to focus. 

The \textit{openness of a network} is characterized by the presence of multiple providers in the market offering customers the opportunity to choose amongst them. Open access network  models separate the ownership of the business actors from the infrastructure layers \ie PIP, NP, SP, with the aim of promoting competition, sharing of the network infrastructure and discouraging vertical integration. 

The following cases can be distinguished although the limits among the respective actors are not always clearcut. 

The different models differ in the functional separation across layers, as recommended by ITU \cite{itu2008}, ranging from vertical integration across all layers in \textit{e, f, g}, partial separation in \textit{a, b, d}, and full functional separation in \textit{c}. The models also differ in terms of alternatives and therefore competition in each layer, except the passive infrastructure, that tend to a single actor in charge of deploying and operating either the backbone or the access area PIP. While all models except \textit{g} offer alternatives in service provision, only\textit{ d, e} provide alternatives in network provision.

Diverse types of local cooperative schemes fit and build on these cases. Municipal networks focus on maximizing the access to connectivity from public (municipal) interest, and they usually rely on public-private partnerships. The service is defined and governed by the public partner but implemented and operated by one or multiple private partners. Typical cases the optical fibre service from Stokab in the Stockholm region, among several other regions in Europe, following the \textit{d} model, or the public WiFi services in most European cities, that can follow any model for service provision, as the public entity just defines, funds and oversees the public service under private operation. Internet eXchange Points (IXP) are physical infrastructure through which Internet service providers (ISPs) and Content Delivery Networks (CDNs) exchange Internet traffic between their networks (autonomous systems). The switching infrastructure is built and managed as a CPR according to Fig.\ref{fig:broadbmodel}, but the governance may range from a centralized \textit{a} to a participatory \textit{CN} model. IXPs and CNs are quite equivalent, the main difference being that IXPs connect larger entities only (wholesale) and CNs focus on individuals and households (retail). However the difference blurs as they expand, with the example of guifi.net that is both a CN and a de-facto regional IXP, or the case of Ninux, that acts like a country IXP of diverse city or regional networks, and connected to the Rome IXP (Namex).
\vspace{1mm}
\subsection{CNs as open access network instances: the commons model}
CNs differ from other models in that there is crowdsourcing in all layers. The community participants contribute and share the passive infrastructure, they also coordinate and operate the active network, and multiple service providers can benefit from that network infrastructure CPR.
Furthermore, CNs embody some key principles \cite{Baig2015150}:

\textbf{Non-discriminatory and open access.} The access is non-discriminatory because any pricing, when practiced, is determined using a cooperative, rather than competitive, model. Typically this results in a cost-oriented model (vs. market-oriented) applying the fair-trade principle for labour pricing \cite{moore2004fair}. It is open because everybody has the right to join the infrastructure.

\textbf{Open participation.} Everybody has the right to join the community. According to roles and interests, several main groups could be identified as stakeholders: i) volunteers interested in aspects such as neutrality, privacy, independence, creativity, innovation, DIY, or protection of consumers’ rights; ii) commercial entities interested in aspects such as demand, service supply, and stability of operation; iii) end users (\ie \textit{customers}), interested in network access and service consumption; and iv) public agencies (local or national), 
interested in 
regulating the participation of society and the usage of public space, and even in satisfying their own telecommunication needs. 
Preserving a balance among these or other stakeholders is desirable, as every group has natural attributions that should not be delegated or undertaken by any other. It is important to clarify that not all stakeholders are present in all CNs. For instance, many CNs object to the participation of commercial entities as this is against their vision and philosophy (e.g. B4RN). 

The model of the CN is based on the concept that the physical and active equipment are used as a Common Pool Resource (CPR). 
Its participants must accept the rules to join the network and must contribute the required infrastructure to do it (routers, links, and servers), but they keep the ownership of hardware they have contributed and the right to withdraw. As a result, the infrastructure is shared and managed collectively, as a collective good.

Comparing the CN commons model with the aforementioned models for open access networks:
\begin{itemize}
\item the CPR (\ie participants of the network, legal entity) replaces the PIP and NP actors;
\item the CPR offers access to private service providers (SPs) but also provides community services (CSs).
\end{itemize}
Cooperation at the network deployment and operation level is crucial but competition in the service provision is encouraged to avoid monopoly situations. 

An example of the commons model in action is provided by the guifi.net CN. The network employs cost sharing and compensation mechanisms in order to facilitate the participation of commercial SPs and operators in the CN. They deliver their services through the network's infrastructure and receive payment from their customers. At the same time, they can contribute infrastructure and invest money to the CPR or compensate the network for using it \cite{Baig:2016:MCN:2940157.2940163}.

\section{CN stakeholders and the sustainability question}\label{sec:CN_sustain_incentives}

Sustainability is a multifaceted concept used to study a variety of systems such as technical, biological and socio-cultural ones. Its precise definition depends on the system of interest. In general, the sustainability challenge consists in understanding the way that a system can smoothly operate in the present and develop in the future. Hence, sustainability is not a specific goal per se but a continuous process to reach a goal. 
Although, originally the term was used in an environmental context (United Nations World Commission on Environment and Development (WCED) 1987), it more recently acquired broader social and economical semantics (World Summit on Social Development, 2005).  

Equally broad is the context of sustainability in the case of community networks, which are by definition complex socio-technical systems. Contrary to the commercial production communication networks, their existence per se is conditioned on the sustained and active participation of all its stakeholders, who contribute resources and generate value for it. Therefore, a sustainable network should first of all ensure that all these actors, primarily end users, but also commercial service providers and public organizations when they are present, have proper commitments and incentives to contribute to the network. This is not a trivial task since the participation of each actor is driven by different types of motives and aspirations, including economical, socio-cultural, and political ones. Hence, the network needs to put in place mechanisms, limits and incentive mechanisms to properly address these aspirations, as in any commons regime~\cite{Ostrom1990}. 

The success of the CN to attract a critical mass of actors also determines the funding alternatives of a CN. A sustainable funding model, which will ensure the network capability to cover its deployment and maintenance expenses, is a crucial parameter for its long-term viability.

We review the practices of different CNs with respect to funding in section \ref{sec:funding}. In the remainder of this section, we describe the broadly varying motives met across and within the different actors in a CN. Then, in section \ref{mechanisms}, we describe how different CNs respond to these motives.

\begin{figure}[tbhp]
\centering
 \includegraphics[width=1\linewidth]{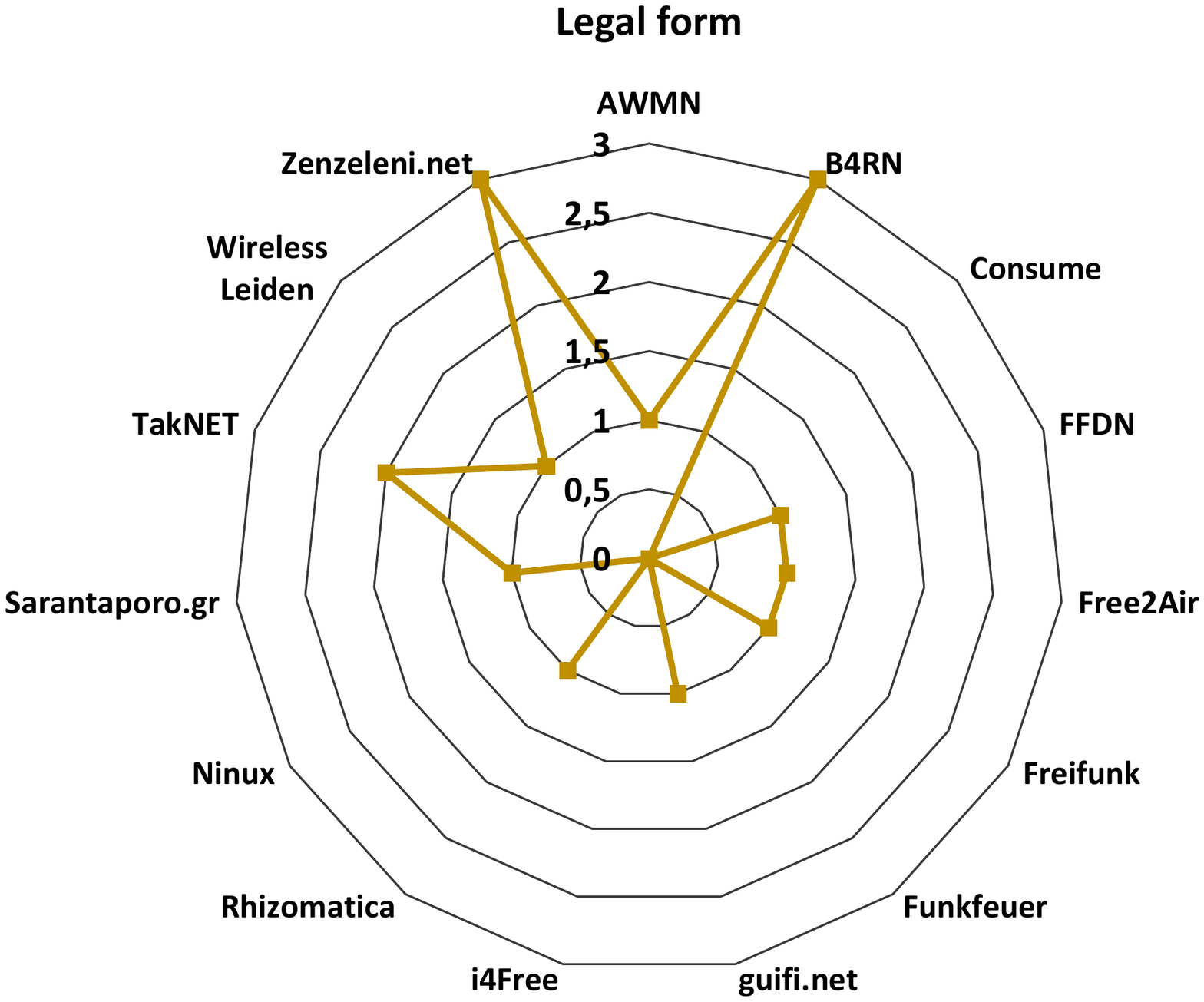}
 \caption{Radar chart with legal forms found in CNs. \textit{CN legal forms}.  0: None, 1: organization (NPO, Foundation), 2: social entrepreneur, 3: operator, ISP.} 
 \label{fig:legal}
\end{figure}

\vspace{1mm}

\subsection{Volunteers}\label{subsec:volunteer}
In the context of CNs, volunteers are the people who initiate the CN project. More often than not, (a subset of) these people take an active role in the network expansion, either through helping with the technical matters and/or organizing informational and training events for potential participants~\cite{DBLP:journals/corr/abs-1207-1031}. 

The volunteer groups usually comprise of people that cumulatively possess knowledge and expertise over a wide set of areas, including technical, legal, and finance matters~\cite{aichele2006wireless}:  
technology enthusiasts, radio amateurs, hackers, (social media) activists, and academics. It is not uncommon for volunteers to create a legal entity (Fig. \ref{fig:legal}) to represent the network to third parties (\ie government, third party organizations, companies, Internet Service Providers (ISPs)). This lets them have a voice and interface with third parties on legal and regulatory matters, but also get involved in financial transactions (\eg collecting user subscriptions, fund raising, purchase of equipment).

Their motives have a strong bias towards political and socio-cultural values and ideals, which is not met in any of the other three stakeholder groups. Experimentation with technology, open software and do-it-yourself (DIY) tools, sensitivity to privacy and network neutrality, the desire to bridge the digital divide, but also commitment to the community spirit and social movement, participatory governance and decision-making, and protection of consumers' rights, count as primary reasons for their involvement in CN initiatives. Economic motivations are much rarer; on the contrary, the members of the volunteers' groups usually end up investing a lot of personal effort, time, and money to the CN initiative, without direct financial return of any kind.
More specifically:

\subsubsection{Socio-Economic motives}
\label{subsec:vol_economic}
Socio-cultural motives often stand behind the original conception and deployment of CNs. 

\textbf{Bridging the digital divide:} The right to (broadband) connectivity is a matter of equal opportunities in the contemporary digital society; and digital illiteracy puts at disadvantage populations deprived of it. The launch of CN initiatives has many times been the response to poor or non-existent access to the Internet and Information and Communication Technology (ICT) services. This is typically the case with remote, sparsely populated rural areas, where commercial operators are reluctant to invest on fixed broadband infrastructure because they do not deem this cost-efficient.

The initial volunteers' group comes typically from local residents suffering the digital divide (as the case is with the B4RN~\cite{NETCOMMONS_D2_2} and guifi networks~\cite{NETCOMMONS_D1_2}). However help may come from outside. In the case of the Sarantaporo.gr network, in Greece, the CN came out of the efforts of a small group of people living in Athens and abroad, with origins from the Sarantaporo area, by the time that no broadband access alternative was available there. 
Likewise, the i4Free network in an island with poor Internet connectivity close to the town of Nafpaktos, in Greece, started from the initiative of a German engineer and professor. He created a small network at his own expenses so that locals could have access to ICT services \cite{NETCOMMONS_D1_2}, \cite{NETCOMMONS_D2_2}, \cite{NETCOMMONS_D2_3}. In a similar scenario, Peter Bloom, an American national, founded Rhizomatica to promote mobile-phone based services in the rural area of Oaxaca, Mexico~\cite{NETCOMMONS_D1_3}.  

\textbf{Economic incentives:} these are in a sense relevant whenever a CN is set up in pursuit of cheaper (affordable) Internet access. In these cases, the underlying idea is how to expand coverage of the service, ensure its sustainability and sovereignty from commercial decisions, and save money with CNs compared to commercial alternatives rather than how to make money out of the CN initiative.

Therefore, in remote, sparsely populated areas, such as the rural areas addressed by the B4RN initiative, the competing alternatives, where they exist, such as satellite or cellular, are typically more expensive and of lower quality. B4RN was conceived also as a way to offer better connections at more affordable prices than its competitors, again in areas where these exist. 

Another case, this time in urban environment, is the Consume network in East London, UK, one of the very first CN initiatives in Europe. James Stevens ran a technology incubation business offering web, live streaming and video distribution services through a leased optic fiber connection. He came up with the idea to connect buildings through wireless mesh links as a way to bypass the expensive license costs and regulatory constraints related to expanding the fiber communication across the buildings.




\subsubsection{Political motives}\label{subsec:vol_political}
Political causes often serve as driving forces for the groups that lead CN initiatives. Such causes often prove to be strong enough to fuel these groups' active involvement with the CN despite the effort, time and money this requires. They include:



\textbf{Openness, net neutrality, and  privacy:} 
These highly controversial issues have served as primary motivations for CN initiatives. The principle of net neutrality dictates that traffic within the network should be treated in an equal manner independently of the type of content or the source. The data communicated across the network is not subject to discrimination. 

A characteristic example of principles underlying the CN initiatives is found in the declaration by the guifi.net Foundation, the volunteers' group that has developed and still operates the guifi CN, in Catalonia, Spain~\cite{Perez2016, barcelo2014bottom, 5514608}:
\begin{itemize}
\item Freedom to use the network, as long as the other users, the contents, and the network itself are respected.
\item Freedom to learn the working details of network elements and the network as a whole. 
\item Freedom to disseminate the knowledge and the spirit of the network.
\item Freedom to offer services and contents. 
\end{itemize}
Moreover, volunteers are often interested in accessing ICT services without having to compromise their privacy. This applies for technology enthusiasts, activists and users in general that wish to protect their private content. 
CNs such as the French FFDN and the German Freifunk declare privacy/anonymity and net neutrality as integral parts of their manifesto and incorporate them in their fundamental operation principles. 

\textbf{Autonomy and alternative communication models:}
These are common motives for the original deployment and subsequent operation of CNs~\cite{lawrence2007wireless}, especially in urban areas, where the digital divide threat is much less pronounced. Community networks such as  Consume\footnote{\url{http://consume.net/}}\footnote{\url{http://wiki.p2pfoundation.net/Consume}} and Free2Air\footnote{\url{http://www.free2air.org/}}\footnote{\url{http://wiki.p2pfoundation.net/Free2Air}} started out representing alternative approaches to the commercial Internet provision, aiming at higher freedom and control over personal communications. In other cases, such as guifi.net, which started as an attempt to bridge the digital divide, such political purposes emerged as an equally strong motivating factor, especially when the number of network connectivity alternatives increased. Rhizomatica founder goals were both to bridge the digital divide and to create an alternate telecommunications network where people could communicate with costs much lower than the existing telecom solutions in the area (where they existed).

\subsubsection{Socio-cultural motives}
Socio-cultural motives 
often stand behind the original conception and deployment of CNs. Among the main ones count: 

\textbf{Experimentation with technology and DIY culture:} Several initiatives are driven by hackers, technology enthusiasts, and academics who enjoy experimenting with network and radio technologies. The involvement within such a community presents them with a unique opportunity to further enhance their technical knowledge and practice it over real networks.

The AWMN, Ninux, and Freifunk CNs were initiated and are still run by network technicians and computer enthusiasts. As such, they have been characterized by a culture of experimentation and improvisation. AWMN and Ninux, in particular are used by their volunteers as testbeds for manufacturing equipment (antennas, feeders) and experimenting with routing protocols and applications. This is evidenced also in the impressive number of native applications and services that were developed for AWMN, without need for public Internet connectivity, including games, libraries, network monitoring tools, DNS solutions, and experimental platforms. Notably, neither AWMN nor Ninux, whose initials stand for "No Internet, Network Under eXperiment", nominally provide Internet access.

\textbf{Community spirit and altruism:} 
Altruism, often coupled with a strong commitment to community ideals serve as important motivations for the active involvement of volunteers' groups in CNs.

Both are strongly evidenced in the B4RN, Sarantaporo.gr and i4Free CN initiatives. 
Community activists have been among the leading figures in B4RN and have set it up as a community benefit society which “can never be bought by a commercial operator and its profits can only be distributed to the community.” 
Likewise, the Sarantaporo.gr non-profit organization involves people who are activists in the area of commons and supporters of community ideals. They place a lot of emphasis on cultivating these ideals in the residents of the area with parallel activities and social events.
Finally, the leading figure behind the i4Free CN, identifies himself as a warm fan of community life and ideals. He has spent enormous amounts of time trying to build a community around the CN through training and educational events, even without much success as he admits \cite{NETCOMMONS_D2_2}.


\subsection{Active participants} 
Even broader is the variety of reasons for the involvement of citizens in a community network. Decisive for many of them is the expectation of available and abundant local connectivity anywhere is needed. Furthermore there is the expectation of cheaper, or even free, Internet access and other services provided by commercial entities. For others, the CN represents a perfect opportunity to acquire new knowledge and experiment with technologies, and/or socialize and become part of a bigger community. Activism in favor of higher autonomy and data privacy are also evidenced as user participation motives, albeit to a smaller extent than in volunteer groups.  

Their levels of participation typical vary a lot within a CN. Some of them are highly active participating in events organized by volunteers or other types of collective activities, sharing their technical experience, developing applications and devoting personal time and efforts to the CN. On the other extreme, a number of users that tends to be the majority in most CNs, set up a node and use the CN to get Internet access or access to local services without further contributing to the activities of the community. 
However, the presence of even these passive users can benefit the network to the extent that others can join the CN through their nodes.


The CN users may pay a \emph{connectivity fee} for being part of the CN or not. These fees contribute to pay the necessary costs to upgrade or maintain the CN infrastructure. Depending on whether they receive some service over the CN, they may pay a \emph{consumption fee} and maintain a contributor, shareholder or customer relationship with the CN, directly or indirectly through paying service fees to a commercial service provider acting as intermediary and value-added reseller. 

\subsubsection{Socio-Economic motives}
Users often expect benefits of economic nature from their participation in a CN, both direct and indirect.

\textbf{Direct economic benefits:} The most usual one is local connectivity or Internet access that is not offered by other providers, or at lower cost than alternative solutions, offered by commercial telecom operators (Table \ref{tab:CN_list}). A characteristic example, Rhizomatica, has managed to reduce costs by 98\% on international (U.S.) calls and 66\% on cellphone calls. Internet connectivity can either be provided by the CNs themselves, which take on the role of alternative Internet service providers (\eg B4RN, Sarantaporo.gr); as an add-on service over the CN by a third party (\eg guifi.net, Rhizomatica); or by CN members who pro bono share their access with other peers (\eg AWMN). 

The collective efforts of the CN participants is often fundamental for expanding the coverage of the network or  lowering the connectivity cost. For instance, B4RN partially crowdsources the cost and effort involved in deploying fiber in rural communities in Northern England. This way, it can offer fiber connectivity and Internet speed in underserved areas and at more favorable prices than alternative commercial solutions. 

A local infrastructure, locally maintained, feeds the local economy, creating paid jobs for the deployment, maintenance, expansion and operation of the network itself and related services over the network (content and services) or enabled by the network (telework, remote assistance, surveillance, sensing).

CNs create the opportunity for local investment. Local can obtain economic benefits from investing in local infrastructures, particularly more durable fibre infrastructures, that can have good returns of investment from usage fees, while also giving indirect economic benefits by increasing the value of households, typically the largest investment of a family.


\textbf{Indirect economic benefits:}
Participation in a CN may incur additional benefits to their users. 
One of them relates to the growth of human capital and another to the added value that the CN generates for businesses and professionals participating in it. Examples from Sarantaporo.gr and AWMN show that young people (in the age of 18-35) view the CNs as a path to information about job and further education opportunities and to business activities developed around the CN \cite{NETCOMMONS_D2_2}. Moreover, in remote rural areas, network access and Internet connectivity can enable professionals to search better markets for their products and cheaper suppliers for their materials (\eg farmers) and small business owners to join the network in the anticipation that visitors appreciate the Internet connectivity feature when choosing where to go (\eg Sarantaporo.gr). 
Underserved communities in terms of connectivity tend to suffer from fragility or lack of other critical infrastructures. The deployment of networking infrastructures creates economies of sharing and bundling, such as improvements in electrification, with the introduction of solar panels, that for instance can enable or improve the quality of night-time lighting and food preservation, which in turn may create economic benefits from trading of these products.

\subsubsection{Political motives}
As seen in section \ref{subsec:vol_political}, many CNs have been initiated under aspirations of privacy, net neutrality, and alternative models of Internet connectivity provision with strong flavor of autonomy and self-organization. 
The ideals underlying the initial development of these CNs are often inherited by subsequent users of the CN. However, these users tend to be a small part of the total CN user population. Typically, the larger the CN grows the harder it becomes to find political causes that unite the whole community behind them.

\textbf{Openness, net neutrality and privacy:}\label{privacy_users} 
The aspects of privacy and neutrality have a strong role in CNs that utilize the Picopeering agreement\footnote{\url{http://www.picopeer.net/PPA-en.shtml}} as a participation/operations framework and are part of the movement for open wireless radio networks\footnote{\url{https://openwireless.org/}.} (\eg Freifunk, guifi.net, Ninux, FFDN). 

The Picopeering agreement is a baseline template formalizing the interaction between two network peers. It caters for a) an agreement on free exchange of data; b) an agreement on providence of open communication by publishing relevant peering information; c) no service level guarantees; d) users' formulations of use policies; and e) local amendments dependent on the will of node owners.

\textbf{Autonomy and self-organization:} 
The participation in CN groups cultivates feelings of autonomy and self-organization. Self organization is practised in the way new users connect to the CN, where they have to rely on their own resources and the voluntary assistance of experienced network members. Being part of an independent network satisfies personal ideological aspirations for self-organized network and autonomous use ~\cite{lawrence2007wireless}. The ability to participate in collective decision making and contribute to an alternative "commons"-based model of ICT access counts itself as a worthy experience for users with strong "commons" ideals. 


\subsubsection{Socio-cultural motives}
A CN is a characteristic example of participatory involvement, where users dedicate their efforts and time to the network~\cite{Vega2014}. A number of services and applications combined with other activities that one way or another revolve around the CN, offer users the opportunity to communicate, educate and entertain themselves, thus further motivating their participation in the network ~\cite{szabo2007wireless},~\cite{pedraza2013community}. \vspace{-3mm}
{ \renewcommand{\arraystretch}{1.2}
\begin{center} 

\begin{table*}[t]
\centering
\resizebox{14.5cm}{!} {
\begin{tabular}{|p{2.1cm}|p{4.2cm}|p{5cm}|}
\hline
\textbf{CN}  & \textbf{Legal form} & \textbf{Funding} \\
\hline
AWMN &    AWMN Foundation  & Members   (individually) \\
  \hline
  B4RN  & Community Benefit Society & Members  \\
  \hline
  Consume &   None & Central actors\\
  \hline 
  FFDN & Non-Profit Organization 
  & Members, Local authorities, Donations  \\ 
  \hline 
  Free2Air 
  & Incorporated Legal Company & Members \\
  \hline
 Freifunk & Non-Profit Organization & Members, Public Institutions 
 \\
    \hline
  Funkfeuer & None &  Members 
  \\ 
  \hline 
  guifi.net &  Guifi.net Foundation  &  Members \\
  \hline 
  i4Free &  None &  Central actor\\
  \hline 
  Ninux &  None & Members   \\
  \hline 
   \multirow{2}{*}{Rhizomatica} & \multirow{2}{*}{Non-Profit Organization} & Members, National and International organizations, Donations \\
  \hline 
  Sarantaporo.gr &  Non-Profit Organization & Members, European Union, Donations \\
  \hline 
  \multirow{2}{*}{TakNET} & \multirow{2}{*}{Social enterprise - Net2home} & Members, Private Insitutions, THNIC Foundation, European Union  \\ 
  \hline 
  
  Wireless Leiden & Non-Profit Organization & Members, Public/Private Institutions \\ 
\hline

Zenzeleni.net & Formal Network/Telecom Operator & Members, Public Institutions \\ 
  \hline 
 
\end{tabular}

}
\caption{CN specific organizational aspects.}
\end{table*}

\label{tab:cn_aspects}
\end{center}
}
\vspace{-5mm}
\textbf{Experimentation and training with ICT:} Technology enthusiasts participate in the network for experimenting with the technology \ie trying software they develop and hacked code, make network speed measurements, play with network mapping and management tools ~\cite{lawrence2007wireless}. Users can acquire new skills about computer and network use, \ie either through self-experimentation or through training by network experts. 

In CNs initiated by volunteers with technical background (\eg AWMN, Ninux, Freifunk), the amount and type of services, applications and self-produced software increased greatly within the community. Besides a variety of network monitoring tools, users can enjoy communication services such as VoIP, online forums, mails, and instant messaging; data exchange services with servers, community clouds and file sharing systems; entertainment services with gaming applications and audio/video broadcasting tools; information and educating services with online seminars, e-learning platforms, and wikis. 

\textbf{Desire for social interaction:} The smooth operation and development of a CN demands  cooperation links at the network infrastructure level but also at the social level. 
In CNs, participants are able to share their ideas and interests, participate in groups, interact and communicate with other network members just like they would in any other online or physical community. Social networking and communication tools raise great interest and remain active even when other tools and services have a drop in their utilization. 

The importance of local relationships in a CN~\cite{kornhybrid} is also evidenced in three independent studies addressing a rural village  in Zambia~\cite{Johnson:2010:IUP:1836001.1836008}, the TakNet CN, in the rural area of northern Thailand~\cite{Lertsinsrubtavee:2015:UIU:2837030.2837033}, as well as Australian and Greek CNs in~\cite{lawrence2007wireless}. In this last study, 91.2\% of the users stated that they enjoyed interacting with the community, 88\% felt that their efforts would be returned by other community members and 80.5\% expressed that the community allowed them to work with people that they could trust and share similar interests. Likewise, in the case of TakNet, 
much of the activity among users of the popular applications such as messaging, email, online social networks and gaming, exhibits a high degree of locality, \ie people use Internet to interact with people within the same CN.

\textbf{Socio-psychological motives:} 
As social motives count socially-aware mechanisms that relate to concepts such as visibility, acknowledgment, social approval, individual privileges and status. This social activity is applied within the networks' technical limits~\cite{Mcdonald02socialissues}. 
 
The ability to compete with other people and satisfy one's self esteem through the involvement in the community, or receive a certain type of credit by others in the community, are motives not as easy to distinguish but still present ~\cite{lawrence2007wireless},~\cite{Lertsinsrubtavee:2015:UIU:2837030.2837033},~\cite{Johnson:2010:IUP:1836001.1836008},~\cite{6673334} and with an impact on network growth and operation~\cite{6979946, 4124126}.

\subsection{Private sector service providers}

Private sector service providers form the stakeholder type that may be less involved in CN initiatives. The term points to companies, ISPs, small businesses or individuals, namely entities that support or use the network to provide some service and get compensated for it. 
These can be a) the professionals that are involved in the installation, operation and maintenance of the CPR network infrastructure, or b) the organizations that provide content or services inside the CN.
At first glance, these entities do over the CN what they do over any other network, \ie provide services where there is demand for them. However, the legal provisions and conditions of running business over the CN may be different given the existence of a CPR infrastructure and the governance and the crowdsourced nature of it.
In fact, the CPR is an enabler of small private sector providers. Since the network commons is a shared resource, that enables these small players to operate and provide services over a larger population, with the economies of scale of cooperative aggregation of CAPEX and OPEX among multiple participants, and the complementarity and opportunities of specialization among them. This also means a lower barrier or entry, with much less initial investment and less risk thanks to the cooperative, and cost oriented model, of the network commons. Therefore the network infrastructure commons becomes a critical resource for the operation and competitiveness of these local private sector service providers. Therefore the common goal would be preserving the commons to enable their specific business models.

The incentives for the participation of private sector service providers in the network are almost always economic.
These actors are interested in profit. The CN provides them with access to potential customers who would otherwise be unreachable. The implementation of their commercial activities depends on the organizational nature of the CN. Guifi.net has set up a framework that enables the participation of private sector in its CN, including maintainers, installers, ISP providers, VoIP providers (Table \ref{tab:guifi_prof}). These entities may sign agreements with the guifi.net Foundation, when the service provision has to do with the sustainability of the CPR. External Over-The-Top (OTT) services, such as Internet VOIP, Video, content providers are left outside. In Rhizomatica, ISPs and VOIP providers are key partners of the organization. Rhizomatica provides the Radio Access Network through which the service providers reach the local communities and their CN users. 
\vspace{-3mm}
{ \renewcommand{\arraystretch}{1.2}
\begin{center} 

\begin{table*}[t]
\centering
\resizebox{16cm}{!} {
\begin{tabular}{|p{3.5cm}|p{10cm}|}
\hline
\textbf{Services} & \textbf{Private sector providers}  \\ \hline

 \multirow{4}{*}{Internet Service Provider (ISP)} &  Adit Slu, Ballus Informatica, Capa8, Cittec, Delanit, Del-Internet Telecom, Ebrecom, Emporda Wifi - Guifi.net a l'Alt Emporda, Girona Fibra, Goufone, Indaleccius Broadcasting, Pangea.org, Priona.net, S.G. Electronics, Steitec-Servei T\`ecnic d'Electronica i Telecomunicacions, Ticae, Xartic 
\\ \hline
  
 \multirow{3}{*}{Mobile Provider}  &  Ballus Informatica, Capa8, Cittec, Delanit, Ebrecom, Emporda Wifi - Guifi.net al Alt Emporda, Girona Fibra, Goufone, Indaleccius Broadcasting, Priona.net, S.G.Electronics, Ticae \\ \hline
  
 \multirow{2}{*}{Surveillance} &   Ballus Informatica S.L., Capa8, Delanit, Ebrecom, Girona Fibra, Goufone, Matwifi, S.G. Electronics, Ticae  \\ \hline 
  
 \multirow{3}{*}{Telephony (VoIP) Provider} &  Ballus Informatica, Capa8, Cittec, Delanit, Del-Internet Telecom, Ebrecom, Emporda Wifi - Guifi.net a l'Alt Emporda, Girona Fibra, Goufone, Indaleccius Broadcasting, Matwifi, Priona.net, S.G. Electronics, Ticae \\ \hline 

TV (IpTv) Provider & Delanit, Del-Internet Telecom, Indaleccius Broadcasting, Priona.net\\ \hline  

 
 \textbf{Agreement types} & \textbf{Service providers} \\ \hline
 
 Economic Activity Agreement & Adit Slu, Asociacion SevillaGuifi, Associacio Guifinet la Bisbal d'Emporda, Ballus Informatica, Capa8, Cittec, Delanit, Del-Internet Telecom, Ebrecom, Emporda Wifi - Guifi.net al Alt Emporda, Girona Fibra, Goufone, Indaleccius Broadcasting, Ion Alejos Garizabal, Maider Likona, Matwifi, Pangea.org, Priona.net, S.G.Electronics, Steitec- Servei Tecnin d'Electronica I Telecomunicacions, Ticae, Xartic \\ \hline
 
 Volunteer Agreement & Cittec, Girona Fibra, Matwifi\\ \hline

\end{tabular}
}
\caption{Private sector service providers in guifi.net and the services they provide.}
\end{table*}
\label{tab:guifi_prof}
\end{center}
}
\vspace{-5mm}
\subsection{Public agencies} 

Public agencies have the natural role of regulating the public space, either for service provision, occupation of public spectrum, public land, but also supporting local development and ensuring access rights to public information and services. 
Public agencies have a responsibility to regulate the deployment and service provision of CNs, as with any other entity performing these activities. Furthermore, they may cooperate with a CN when the mission of both align. They may contribute to its deployment and growth through funding the initiative, sponsoring network equipment, consuming CN services, facilitating its expansion and growth or by permitting the use of public space and resources by a CN. In Catalonia, the Foundation operating guifi.net has developed the Universal format~\cite{universal17}, a template municipal ordinance, that allows municipalities to regulate public, commercial and community entities to deploy shared infrastructures in public space. Under these principles, several local authorities have allowed guifi.net groups to dig public space and lay down fiber for expanding the network. In several German cities, Freifunk is given the permission to set up antennas and equipment in the roof top of churches, Town Halls, or other public buildings. 

Quite often other types of public agencies get involved in the network. 
Sarantaporo.gr has received network equipment for the initial deployment by the Greek Foundation for open-source software and Internet connectivity from the regional University of Applied Sciences. TakNet received financial support from the Thai Network Information Centre Foundation and initial equipment donation and support from the Network Startup Resource Centre. 

Depending on their level of participation public agencies 
can sign collaboration agreements with the legal entity of the CN and contribute economic or infrastructure resources with or without compensation.

\subsubsection{Socio-economic motives} 
The participation of public agencies in a CN initiative can also have an economic motivation. In the case of guifi.net public agencies can fund the network expansion through purchase of equipment in return for complimentary added value services over the CN. Public agencies may be interested in the added value of purchasing connectivity services from a CPR infrastructure, as while being competitive in price, can amplify the spill-over effects in the local economy, and contribute to socio-economic development. However, public entities may also be tempted to put obstacles as a result of the influence and pressure of traditional large telecom companies, with more taxation than large telecom or Internet players that may enjoy unfair tax benefits.

\subsubsection{Political motives}

The participation of public agencies in a CN often comes as a result of high-level policies against the digital divide, to increase the offer or lower the costs of local connectivity, and in favor of equal opportunities in the digital economy and society. 

\subsubsection{Socio-cultural motives} Public agencies may also support CNs because they acknowledge their long-term potential to strengthen the community links, raise awareness for issues concerning the local societies and favor the engagement of citizens with the commons. On the polar opposite and more opportunistic note, local administrations (such as municipalities) can advertise the provision of network services as a political achievement that increases their re-election chances. 

\section{Funding Sources for CNs}\label{sec:funding}

CNs use one or more of the following ways (Table \ref{tab:cn_aspects}) to fund their activities~\cite{NETCOMMONS_D2_4}:
\subsection{Member subscriptions and contributions in kind} 
This is the most common funding model for CNs. In this case, the members of the CN contribute network equipment and time/effort to the network growth and maintenance. In the case of the BARN network, which provides fibre connectivity, members even contributed digging effort. In most cases, the CN users pay a monthly/annual subscription fee for the CN needs. Several CNs such as AWMN in Greece, Ninux.net in Italy, BARN in UK, and Freifunk.net in Germany, have managed to scale significantly this way.

Despite its simplicity, the model has several variations. Subscriptions may be mandatory or voluntary; or they may serve as a prerequisite for participation in decision-making bodies and voting rights. In the case of the Sarantaporo.gr, it is villages under the network coverage, rather than individual CN users, that are charged with a fee. How each village will split the cost among local users is left to the the CN participants in that specific village to define. 

What the CN users get in return for their subscriptions is closely related to the way the CN organizes itself and positions in the telecommunications arena. 
For example, B4RN operates as a community benefit society, which provides Internet service to its subscribers. The subscription model is composed of a connectivity fee and different service fees for different types of users. On a similar note, Zenzeleni.net operates as a cooperative telecommunications operator providing voice and data services to its customers. TakNet has developed a social enterprise called Net2Home. Users have to pay monthly fees that are used for covering fiber (to the network operator), maintenance, equipment installation, technical online support, network management and monitoring costs. Rhizomatica helps communities in Mexico build their networks receives a flat rate for equipment installation and community member training as well as a percentage of monthly subscription, advisory and technical services fees. Finally, but far more rarely, a CN may operate as a for-profit company. Some of the FFDN networks in France are commercial networks that indeed rely on policies such as standard pay-per-use contracts and added value services to customers outside the CN. However, in contrast with traditional commercial companies that extract profit from customers and locals to retribute investors, CNs reinvest the profits in the CPR infrastructure.


\subsection{Donations from supporters} 
Community Networks are often financed through crowd-funding projects or direct, regular or one-time, donations.
In some CNs, citizens can invest in the infrastructure, either for a specific reason such as crowdfunding the construction or improvement of a critical link that affects the user (typical in guifi.net), or generic, through community shares to expand the local network or even his home access (in B4RN). These investments can generate tax returns (guifi.net Foundation or B4RN). In B4RN this investment also generates (3\%) interest after the third year.
In developing areas or in disaster situations external donors can contribute funds such as in Zenzeleni (ZA), Rhizomatica and Nepal Wireless (NP).
This funding source typically complements other funding sources since it rarely suffices to cover the CN's funding needs.

\subsection{Support from public agencies and institutions} There are cases, where CN initiatives have got generous support from public funds (cash or in kind). Municipalities and local authorities emerge as main actors in this respect. The synergy of commons/public service with civil society/municipality can limit the survival concerns of CNs as far as one finds sustainable models that motivate their cooperation~\cite{RePEc:ehl:lserod:29461}, \cite{powell2006going},\cite{powell2008wifi}.

One such case is the Sarantaporo.gr CN, which set up its first nodes with hardware and equipment received from the Greek Free/Open source Software Society (GFOSS); and, later, expanded the CN through funding by the CONFINE project~\cite{braem2013case}, funded by the European Commission.
Likewise, in the case of Freifunk, the support from public authorities was expressed through making available public buildings such as churches or Town Halls for placing and storing the network's equipment (\eg antennas).

Sometimes, the support may be expressed in more indirect, yet equally significant, ways such as giving proper attention to CNs in regulatory actions. The guifi.net Foundation has developed a cooperative infrastructure sharing model (the “Universal” deployment model) that develops over the Directive 2014/61/CE on broadband cost reduction of the EU\footnote{\url{https://ec.europa.eu/digital-single-market/en/news/factsheets-directive-201461ce-broadband-cost-reduction}} and the infrastructure sharing concept of the ITU\footnote{\url{https://www.itu.int/ITU-D /treg/publications/Trends08_exec_A5-e.pdf}}. The model prescribes how municipalities and counties can regulate the use of public space by private, government and civil society in a sustainable manner~\cite{universal17}. Rhizomatica has expressed interest in following a compensation system such as the one used in guifi.net~\cite{NETCOMMONS_D1_3}. 
\vspace{-1.2mm}

\begin{figure}[tbhp]
\centering
 \includegraphics[width=0.9\linewidth]{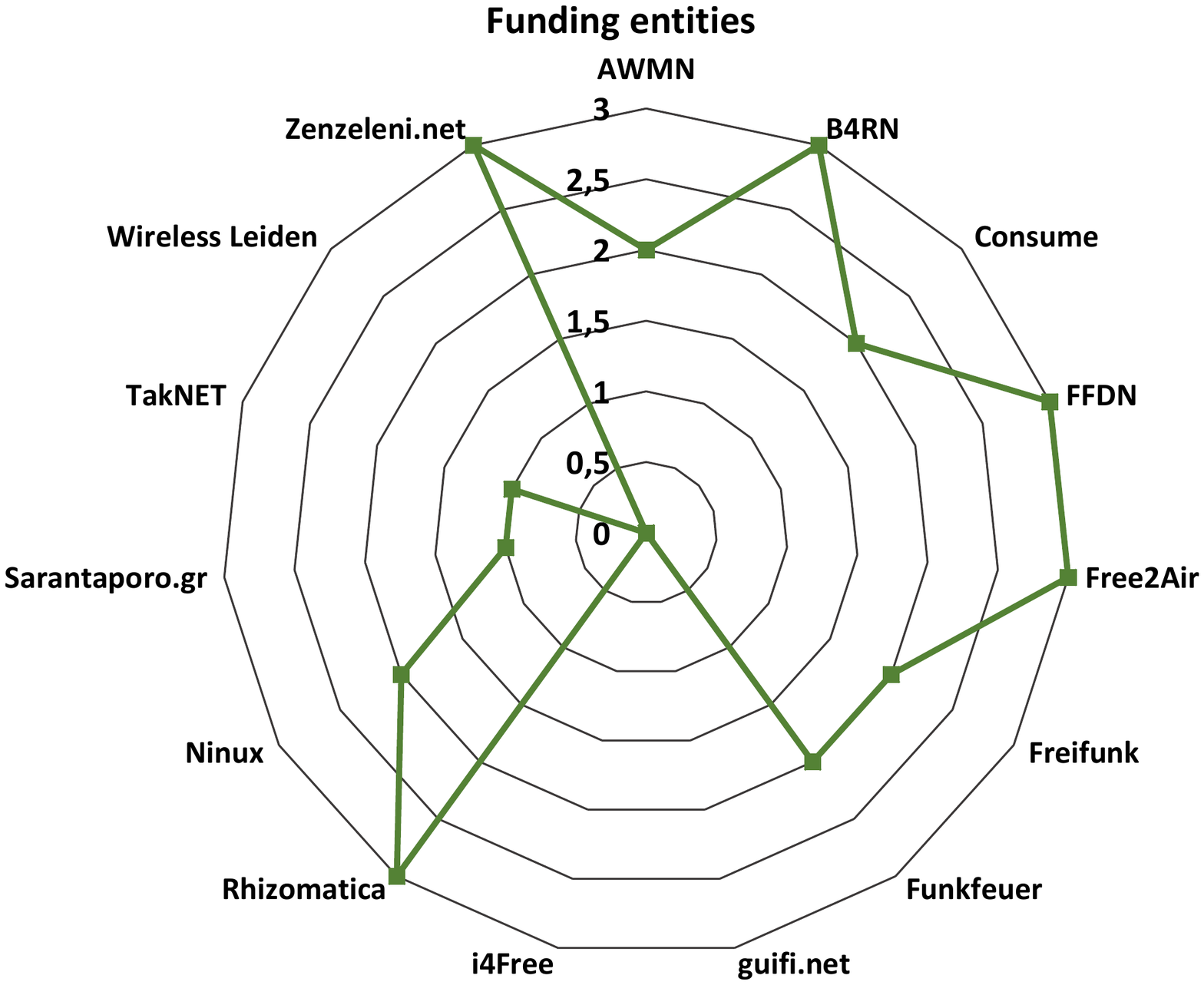}
 \caption{Radar chart with CN funding entities in a 0-3 scale. \textit{CN funding sources}. 0: mainly private entities' involvement, 1: mainly public agencies' involvement, small scale member contribution 2: mainly member contribution (donations, non regular fees), 3: member contribution only (regular fees).} 
 \label{fig:chart2}
\end{figure}
\vspace{-1.2mm}
\subsection{Funding from private sector through commons-based policies}
In the case of guifi.net, CNs have come up with 
unique innovative models combining voluntary and professional services into a commons-based approach. Commercial service providers offer services over the CN and charge the CN users as typical customers, but also subsidize the CN growth and maintenance subscribing to the commons policies. This way, the CN maintains its non-profit orientation and pursues its sustainability through synergies with entities undertaking commercial for-profit activities~\cite{Baig:2016:MCN:2940157.2940163}. 
\vspace{0.5mm}

When assessing the strengths and weaknesses of the four categories of funding sources, the following remarks are due: 
\begin{itemize}
\item Some sources (\ie donations, voluntary contributions \etc) are one way or another not guaranteed and they make long-term strategic planning difficult. They could also lead to disagreements and conflicts between CN members concerning their distribution inside the network, especially if there are not well-defined decision-making processes.
\item Unless something dramatically changes on the regulation side, the support of public authorities for CNs cannot be taken for granted. BARN is one CN instance that tried to access national funding without success (their bid for the funding was eventually withdrawn). In the case of guifi.net, the municipality of Barcelona is reluctant to provide the CN with access to the city wi-fi and fiber infrastructure. In general, CNs tend to view access to local, national or European funds too difficult as well as demanding, uncertain, and bureaucratic.
\item The dominant view across CN initiatives is that the funding from own resources is the most reliable and favorable option. BARN and Freifunk, two of the three networks in Europe that have managed to scale in the order of tens of thousands of nodes, have followed this approach.
\item Trying to put commercial service providers in the loop while preserving the CN ideals, as guifi.net does, definitely represents an innovative approach. The success it experiences in the case of the guifi network renders it a valid funding model alternative.
\end{itemize}
\vspace{0.3cm}
Interestingly, only guifi.net so far has managed to involve in its funding model all possible actors (end users/members, private sector and public authorities). Striking the right balance between the roles and contribution modes of these three parts may prove the key towards the economic sustainability of CN initiatives.

Fig. \ref{fig:chart2} summarizes the funding dependencies on different sources for the 15 CNs. Member contributions, public or private institutions, public authorities, contests and funding projects are met in different scales within each CN and provide part (or all) of the network's resources.
Some CNs operate through regular economic contributions of their members in commercialized subscription models (B4RN, Zenzeleni.net, FFDN, Rhizomatica) while others adhere to non regular fees usually gathered in the form of donations by their members (AWMN, Freifunk, Funkfeuer). In cases where the contributions by a CN's own members are not systematic, public fundings (Sarantaporo.gr, TakNET) pose a significant aid and private entity involvement contributes to the economic activity of the CN (i4Free, guifi.net, Wireless Leiden). The involvement of private entities varies among a single person's funding met in i4Free, ISP's offering their services and contributing to the ecosystem of the CN in guifi.net, or private and public funding by institutions in Wireless Leiden.

\section{Incentive Mechanisms in CNs} \label{mechanisms}

 
To ensure a sustainable presence, CNs have put in place diverse incentive mechanisms. As with other types of commons~\cite{Ostrom1990}, the main purpose of these mechanisms is to limit, encourage and fuel the original motives for participation of all types of actors. They also aim to prevent phenomena and conditions that might weaken the original motivation of actors. Such phenomena include mainly:

\emph{Free riding and selfish behaviors:} many users are solely interested in enjoying network connectivity without themselves contributing adequate or any resources to the CN. Such behaviors can easily lead to the depletion of network resources and CN degradation. 
Mechanisms for organizing and ensuring users sustained contributions and distributing effort across them are of significant importance.

\emph{Unclear CN legal status:} CN actors (users or private sector entities) may be deterred from joining the network and participating in its activities if its legal status is not clear. Well established operational and participation rules can alleviate such effects.

In what follows, we review incentive mechanisms that are either in place in different CNs or have been proposed, without (yet) finding a path to implementation, in the literature. In the latter context, we also review mechanisms that have been proposed for \emph{similar} systems such as wireless ad-hoc networks, P2P systems, and virtual online communities. These systems display inherent structural similarities with CNs in that they also depend on the collective effort and cooperation of their participants to fulfill their tasks: forward and route data in wireless ad hoc networks, disseminate files and other data in P2P systems, share effort and data in virtual online communities.

\begin{figure}[tbhp]
\centering
 \includegraphics[width=1\linewidth]{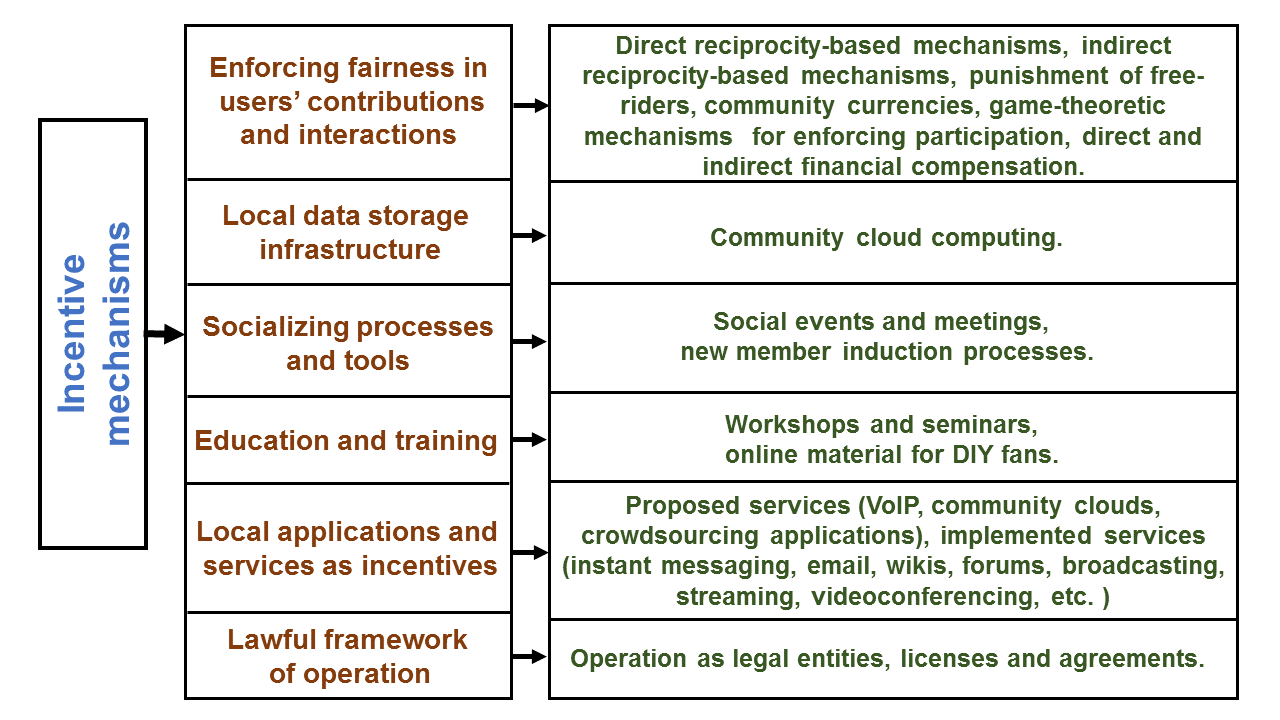}
 \caption{Categories of incentive mechanisms used in CNs.}
\label{fig:mechanisms}
\end{figure}

The different incentive mechanisms aiming to motivate the participation in CNs and strengthen their sustainability are grouped into six categories (Fig. \ref{fig:mechanisms}).

\vspace{-1.5mm}
\subsection{Enforcing fairness in users' contributions and interactions}\label{s:enforcing}

Despite the direct threat that free riding phenomena pose to the network's long-term sustainability, actual prevention countermeasures are not that widespread in most CNs, with the notable exception of guifi.net~\cite{Baig2015150}. Interestingly, a quite broad range of solutions have been proposed in the literature, either in the specific context of CNs or that of similar systems (wireless ad-hoc, P2P, and online virtual communities) \cite{GhLo-2011, CiLo-2011}.

\subsubsection{Direct reciprocity-based mechanisms}
Reciprocity is a broad term that incorporates the notion of human cooperation in different interaction scenarios~\cite{nowak2006five}. \textit{Direct reciprocity} keeps records of the interaction of two specific individuals so that the accounts are settled between those two. The "tit-for-tat" manner of connecting to wireless CNs is quite common practice between their members. For a node to connect to a CN, there must be another node to which the connection is directed. In many cases, the reciprocal sharing obligations stemming from the participation in the CN, are explicitly described in licenses such as the Wireless Commons License (WCL)\cite{Baig2015150}\footnote{\url{http://wiki.p2pfoundation.net/Wireless_Commons_License}} defined in terms of neutrality and general reciprocation. 

Direct reciprocity mechanisms can be described in various contexts such as in sharing network connectivity or storage and computing resources. The compensation tables in guifi.net is a key resource to ensure the economic sustainability of the network, ensuring a cooperative and cost-oriented model to share the recurring costs and balance investment, maintenance and consumption~\cite{Baig:2016:MCN:2940157.2940163}. 

In terms of proposals, connectivity sharing is the objective studied in \cite{4146973}. A reciprocity algorithm, coupled with the P2PWNC protocol in~\cite{efstathiou2006practical}, keeps account of the services each participant provides and consumes via technical receipts. This way, it keeps a balance between   the amount of traffic users transfer and that they relay on behalf of others. The model considers the provision of Internet access through the APs of a wireless CN. Participants are divided into teams that manage their own APs and consume/contribute traffic of/to another AP. 

Reciprocity-based mechanisms for sharing storage and computing resources are reported in~\cite{6583482},~\cite{buyukcsahin2013incentive} and~\cite{Vega:2013:SHR:2595405.2595411}. In~\cite{6583482} and~\cite{buyukcsahin2013incentive}, the reciprocity-based mechanism is implemented over a Community Cloud made out of shared computational resources of the network members and is based on records of participants' efforts. Results indicate that the most suitable structure for community clouds should distinguish between ordinary nodes that possess cloud resources and super nodes that are responsible for the management of resource sharing. In~\cite{Vega:2013:SHR:2595405.2595411}, mobile devices used for computing, borrow CPU slots in a reciprocal manner. It is suggested that the heterogeneity in the amount of available resources may not be beneficial for participants with large-scale resources. 

\subsubsection{Indirect reciprocity-based mechanisms}
The concept of direct reciprocity readily expands to that of \textit{indirect reciprocity}, which is essentially realized by reputation mechanisms. Indirect reciprocity does not consider two specific individuals (like direct reciprocity) but rather asymmetric random exchanges based on the reputation scores of each individual node. 
Key issues in building reputation mechanisms \cite{4468733}, involve keeping past behavior records (as node reputation is partially built over time), carefully evaluating all of the acquired information and distinguishing between old data vs recently gathered ones. Among other challenges, reputation-based systems have to face the impact of liars on peer reputation \ie nodes giving unreliable information about other nodes. The system should be able to yield immediate response to known misbehaving nodes by drawing from past information.

The guifi.net classification of suppliers\footnote{\url{https://guifi.net/en/node/3671/suppliers}} (professionals, volunteers) provides a public ranked list according to reputation of professionals and volunteers available for a range of tasks. The list is based on the certification of their abilities based on actual deployments or training courses.
 
Reputation mechanisms have been proposed for P2P systems and wireless ad-hoc networks. In \cite{4343463}, such a mechanism is developed to build a reputation score for P2P system participants. Each peer is described based on how much service (bandwidth, computation) it provides and consumes. Peers are encouraged to collaborate with each other and receive an increase in their reputation metrics. The mechanism successfully results in peers making coalitions that eventually work to their benefit. In a similar rationale for routing in mobile ad-hoc networks (MANETs), the reputation technique aims at isolating non-cooperative node behavior using the Confidant protocol. The tamper-proof hardware, which is embedded in nodes, keeps account of their \emph{virtual credit} collected as they contribute in packet forwarding. The reputation mechanism in~\cite{Michiardi:2002:CCR:647802.737297} keeps records of the collaboration activities of nodes in the MANET and builds a reputation score for each node, based on monitored collaboration data and information input from other nodes.

\subsubsection{Punishment of free-riders}
Free riding is a quite common problem in commons, experienced in various forms by each network type. The design of long-enduring CPR institutions~\cite{Ostrom1990} requires graduated sanctions for appropriators who do not respect community rules. 

This implies defining the ``boundaries'', determined by the community license and agreements, and requires effective conflict resolution methods that may include sanctions \cite{Baig2015150}.
The conflicts resolution system in guifi.net provides a systematic and clear procedure for resolution of conflicts with  participants  that negatively affect the common infrastructure resource, with a scale of graduated sanctions.  It consists of three stages —-conciliation,  mediation,  and arbitration—- all of them driven by a lawyer chosen from a set of volunteers.  This has been found critical to keep the infrastructure and the project itself operational. 

In multi-hop wireless networks, consumption of bandwidth and energy serve as the main motivations for nodes' free riding behavior. Nodes enjoy packet forwarding of their own packets by other nodes but defer, either deterministically or probabilistically, from forwarding the packet of other nodes. Detection and punishment of suspected free-riding nodes are the two basic steps suggested for dealing with this phenomenon in the corresponding literature.

In the generic setting in \cite{8bd91e0a0a71406681d682e5d29b4bbd}, it is suggested that free riding should be confronted using exclusion of peers from a group as a plausible threat. Misbehaving nodes are detected through reputation protocols and excluded from the network or community. Detection of selfish behaviors of mesh routing nodes is carried out in~\cite{Martignon:2009:FDS:1641944.1641958} with a trust-based mechanism. The mechanism can be developed based on the combined observations of neighbor (and other) nodes of the CN such as in KDet \cite{lopez2015kdet}. The Catch protocol in~\cite{Mahajan:2005:SCM:1251203.1251220}, tries to limit the free riding problem in multi-hop wireless networks while preserving anonymity. The adopted technique uses anonymous messages and statistical tests to detect the selfishly behaving nodes and isolate them. It relies on the assumption that free-riding does not appear in the initial stages of the network deployment but later, as the number of peers starts to grow. The corresponding example in CNs reflects the fact that the initial members, \ie volunteers, create the CNs based on certain principles and knowledge that are not compatible with free riding practice. Members that join the network in subsequent stages, \ie users, are often not acquainted with these principles and the importance of complying to them.

\subsubsection{Direct and indirect financial compensation}
\label{subsec:compensation}
This type of mechanisms aim to support CNs' economic sustainability. Guifi.net, a representative example of this category, involves private sector actors that provide commercial services in the CN. To this end, it has set forth additional mechanisms for compensating contributions of different stakeholders \ie compensation system, provision of donation certificates~\cite{Baig:2016:MCN:2940157.2940163}.

The compensation system aims at settling imbalances between network usage and contributions (CAPEX or OPEX). It is a way for participating entities to share network costs while acquiring network resources. Private sector service providers may assume the roles of operators that contribute to the network and consume its resources, investors that only contribute, and pure operators that only consume network resources. Operators can contribute either to the deployment of the infrastructure or to its maintenance. 

The provision of donation certificates that are amenable to tax deductions, is a way of acquiring indirect benefits for contributing to a commons infrastructure. Users who pay commercial service providers for service provision, can have some tax deduction benefits as well according to the Spanish legislation and regulation authorities. 

Other mechanisms explored in the literature but not yet validated in CNs are the following.

\subsubsection{Community currencies}
\label{subsec:currencies}

The design of community currencies is a way to enforce reciprocity and balance the contributions of nodes to the network. As long as the cost/value of nodes' contribution can be quantified, community currencies can ease the exchange of a wider set of services between CN members and users of a CN and properly reward voluntary activities. At the same time, community currencies are themselves collaborative activities that increase the community spirit and strengthen the intrinsic motivations for participating in a CN. In fact, the smooth operation of a community currency depends heavily on building trust between community members both to accept and use the corresponding currency but also to be able to provide risk-free credits that are very important for the required flow of currency. This trust is a very important asset that can play a key role in the initial birth and sustainable operation of CNs. For the same reason (existing trust and community values), the existence and operation of a CN eases the launch of a community currency. The development of community currencies for CNs is yet at an initial stage but they pose a promising mechanism that exhibits a complex bidirectional relation with CNs~\cite{NETCOMMONS_D2_4}.


\subsubsection{Other game-theoretic mechanisms for enforcing participation} Participant's motives for contributing in CNs can be enhanced by game-theoretic and mechanism design approaches. An incentive mechanism based on a Stackelberg game is provided in~\cite{Biczok:2011:IGW:1942329.1942565}. The objective is to stimulate user and ISP provider participation in a hypothesis of a global CN where the participating entities (users and ISPs) interact with an intermediate entity, \ie the community provider or mediator. 

Due to the cooperative nature of CNs, participation of peers often needs to be combined with mutual cooperation \ie forwarding packets, amongst them. While some works use reputation-based mechanisms there are others that prefer credit as a plausible economic incentive to sustain participation. The works in~\cite{Feldman:2004:RIT:988772.988788},~\cite{1610590},~\cite{zhong2003sprite} and~\cite{Zhong2007} tackle this objective in different types of systems \ie P2P, static or mobile ad-hoc systems. 

In a P2P network setting~\cite{Feldman:2004:RIT:988772.988788}, the prisoner's dilemma is chosen to design incentive techniques and deal with challenges such as large populations with small lifetime, asymmetry of interest in participation and multiple peer identities. In order to enhance cooperation and avoid false identities and hijacking, the mechanism proposes to keep records of peer interaction and use them to build reputation metrics. In another approach, the work in~\cite{1610590} uses game theory techniques to enhance cooperation in static ad-hoc networks and suggests that the most effective incentivizing structure is one that combines actual incentive mechanisms \ie actual credits as reputation systems or virtual currencies, with mechanisms that target players' self interest and enjoyment. A Video on Demand service on wireless ad hoc systems is the setting for the Stackelberg game presented in~\cite{6260452}. In order to promote cooperation among participants \ie upload and forward data, the content provider offers them rewards which vary across actual payment, virtual credit or reputation points. A software protocol in~\cite{zhong2003sprite} combined with a game-theoretic aspect is used to stimulate cooperation among selfish nodes in mobile ad-hoc networks. A cheat-proof and credit-based mechanism determines node rewards and costs which are utilized for packet forwarding and route discovery. 

\subsection{Community cloud infrastructure}
\label{sec:privacy} 
\label{sec:cloud}

CN services and applications that store data or process locally can serve as privacy-related incentive mechanisms for CN participants avoiding the exposure to not well understood and often privacy-unfriendly practices of commercial data storage solutions. More often than not, such services involve the deployment of distributed cloud solutions that are deployed locally across the CN nodes, that process and store users' data without dependence on external cloud services.

A proposition to extend CN resource sharing beyond bandwidth resources to computing resources can be found and discussed in~\cite{6673334}. Cloud computing infrastructures can be developed in various ways but face severe challenges due to the nature of CNs \ie hardware and software diversity with various options for inexpensive material, decentralized management where users contribute and manage their own resources and rapid changes in the number of contributing nodes. The idea of developing a distributed Community Cloud that follows the topology of CNs is proposed in~\cite{6583482}. The goal is to regulate consumption and contribution of participant resources in the community cloud in accordance to one's level of contribution. They present an effort-based mechanism for stimulating node participation in resource sharing. Nodes are incentivized using rewards that depend upon their contribution, \ie effort to the local cloud system. A Community Cloud can also be used in conjunction with Grid Computing techniques~\cite{Marinos:2009:CCC:1695659.1695704}. The Community Cloud uses the spare resources of network nodes while considering environmental sustainability and self-management and replaces vendor clouds with full access to users' resources.
 
\subsection{Socializing processes and tools}
\label{s:socializing}

CNs have developed a great variety of mechanisms to promote participation, interaction and knowledge dissemination among CN members \ie social events, meetings, new member induction process. These mechanisms serve as a "social" incentive to encourage active involvement and engage new and old members to CN processes and operation.

\subsubsection{Social events and meetings}
Large- and small-scale CNs organize gatherings and events to discuss not only CN organizational matters but also strengthen the bonds of community members through social activities.
Face to face meetings are common practices. Depending on their morphology \ie a single network or network of networks, CN members have meetings weekly, monthly or annually. CNs which are composed of smaller networks (guifi.net, Ninux, Freifunk), tend to have weekly or monthly face to face meeting at the local networks and an annual global meeting to get together and discuss the issues arising from the operation of the entire network. Other CNs like AMWN, schedule frequent meetings (\ie General Assembly) when important organizational matters are up for discussion. 

\subsubsection{New member induction processes}

Depending on the mentality and philosophy of the particular CN, interaction with network members is a natural prerequisite for a newcomer's access to the network. The way that this interaction is later on retained, is possible to determine their individual participation level. 
For example in AWMN or guifi.net, new participants are urged to register and communicate with nodes of physical proximity to them. After communicating with the node owners, they are able to receive advice about the equipment they need and acquire assistance from existing members in setting up their own nodes and joining the network. Many node owners provide public contact information for others to contact them. In cases, where actual interaction with node owners is not possible or for complementary assistance, users can register to the website and post their questions in the CN's forum.
\vspace{1mm}
\subsection{Education and training practices}
\label{s:education}

Education and training of CN members is an important aspect of CNs, addressing their members desire for acquiring new skills and learn more about networking and radio technologies. Seminars, workshops and online manuals are the main deliverables of this line of effort, invested typically by members of the volunteers' group but also by other CN members.

\subsubsection{Workshops and seminars}
Several workshop and seminar events are organized by existing CNs (AWMN, Sarantaporo.gr, guifi.net). Experienced members share their knowledge with new members, exchange ideas and present available technical solutions. Guifi.net is quite active in organizing workshops and training seminars \ie guifi labs\footnote{\url{http://www.guifiraval.net/}} \footnote{\url{https://guifi.net/en/event}}, the SAX\footnote{\url{ https://sax2016.guifi.net}}, or supports related events FOSDEM\footnote{Free and Open Source Software Developers' European Meeting: \url{https://en.wikipedia.org/wiki/FOSDEM}}, the Dynamic Coalition on Community Connectivity (DC3)\footnote{\url{https://www.intgovforum.org/cms/175-igf-2015/3014-dynamic-coalition-on-community-connectivity-dc3}}.  

AWMN workshops aim at enhancing members' technical skills by disseminating knowledge and technical expertise, interacting with people that have the same interests, strengthening the bonds within the community and new member training. In a different approach, Sarantaporo.gr workshops are more focused to the broader community of locals (with or without technical expertise), inform people about the operation of the network and share knowledge over the wireless networking principles and the development of community networks. 

\subsubsection{Online material for DIY fans}
CNs invest effort to derive manuals and how-to documents so that users can learn more about technical matters and be able to set up their own nodes. Freifunk, Ninux, AWMN, guifi.net follow this practice and develop guides that provide technical instructions on actions and requirements of setting up nodes, FAQs and other useful information. Participants are encouraged to self-educate and "take matters into their own hands" instead of relying to "experts" and behaving as consumers of service. In cases, where online material is not enough they can always get advice in CN forums, or retrieve contact info of node owners.
\vspace{-0.2cm}
{  \renewcommand{\arraystretch}{1.1}
\begin{center} 

\begin{table*}[t]
\centering
\resizebox{14cm}{!} {
\begin{tabular}{|l|c|c|c|c|}
\hline
\textbf{Mechanisms} & \textbf{Volunteers} & \textbf{Users} & \textbf{Private sector service providers} & \textbf{Public agencies} \\
\hline
 Direct reciprocity &   &  x   &   & \\
  \hline
  Indirect reciprocity  &  & x &  & \\
  \hline
  Punishment of free-riders & x &   &  & \\
  \hline 
  Community currencies &  & x & x & \\ 
  \hline 
  Game-theoretic &  & x & x & \\
  \hline
  Financial compensation &   &   & x &  \\
    \hline
  Local data storage infrastructure &  & x &  &  \\ 
  \hline 
  Social events and meetings & x & x  &   &\\
  \hline 
  New member induction processes &   & x &   & \\
  \hline 
  Workshops and seminars &   & x &   &  \\
  \hline 
  Online material for DIY fans &   & x &  & \\
  \hline 
  Local applications and services &  & x &  & \\ 
  \hline 
  Operation as legal entities &  & x & x & x\\ 
  \hline 
  Licenses and Agreements &   & x & x & \\ 
    \hline 
\end{tabular}
}
\caption{Incentive mechanisms and relevance to the CN stakeholders.}
\end{table*}

\label{tab:mechanism_stakeholder}
\end{center}
}
\vspace{-4mm}
\subsection{Local applications and services as incentives}\label{s:local}

The applications running over the network can themselves be considered as mechanisms motivating persons to join the network\footnote{There are arguments both in favour of the importance of local services in CNs~\cite{antoniadis2016}, but also doubts that local services can make an impact on CNs considering that public Internet covers any application needs on the side of the user ~\cite{NETCOMMONS_D2_2}.}. Such services range from network connectivity to communication and entertainment. 

\subsubsection{Proposed services} The CN literature has shown special interest in services such as VoIP, community clouds and crowdsourcing applications.
Trusted VoIP service for nomadic users in wireless network scenarios is the subject of~\cite{efstathiou2006building},~\cite{Frangoudis2014330} and~\cite{6979951}. Building upon an existing scheme, the Peer-to-peer Wireless Network Confederation (P2PWNC), the work in~\cite{efstathiou2006building}, develops a VoIP scheme utilizing residential Wireless LAN access points (producers of bandwidth) for nomadic users (consumers of bandwidth) as a low cost alternative to traditional GSM telephony. In another generic setting in~\cite{Frangoudis2014330} and~\cite{6979951}, authors experiment with VoIP services for nomadic users using community-based Internet access and identify their perceived challenges (trust on nodes, data privacy and unspecified conditions of the wireless environment) and performance limits (capacity, service quality, security).

Cloud services have also attracted a lot of attention as fundamental privacy enablers, \ie for storing the data of the CN locally, without needing to interact with the public Internet. A detailed discussion on clouds can be found in section \ref{sec:privacy}.

Crowdsourcing applications have the potential to match very well the participatory nature of wireless community networks, \ie participatory networking~\cite{Vega2014} and the strong community-oriented social structure met in most developing regions. In the crowdsourcing paradigm, individual users solicit information, content or service from groups of people. The community dimension only strengthens the case for such applications since the community bonds serve as additional socio-psychological incentives for the active participation and contributions of end users. The common resources shared by the members can serve as the media where users (mobile or not) connect to post tasks or get informed about available task announcements. Users receive explicit rewards such as monetary payments, virtual credits of services that match the services they offer~\cite{6275807}. 

\subsubsection{Implemented services}
Certain CNs have implemented a broad variety of services and applications, while others are at a more initial stage of service and application provision. In CNs like Sarantaporo.gr and i4Free, the main service of interest is Internet access. Yet, Internet access is not always on offer by the CN: Ninux does not provide any Internet service at all; guifi.net offers the ability only through private Internet service providers operating over the CN; and, in other networks, such as the AWMN, members occasionally share their Internet connections with other users through APs. 

Networks built by people with technological background tend to elaborate more on the provision of non-professional services. Tools for communication such as chat, email servers, mailing lists, wikis, forums, data exchange, entertainment like broadcast radios, podcasts and streaming are common services found in most CNs (AWMN, Ninux, Freifunk, guifi.net). AWMN and Ninux users have also access to VoIP and chats, guifi.net users to videoconferencing, AWMN and guifi.net users to local clouds and FFDN and Freifunk and AWMN users to collaborative writing tools. Apart from the basic services used in most CNs, there are also several CN specific ones \ie multi-player gaming, broadcasting, live streaming, e-learning, local search engines (Quicksearch, Wahoo, Woogle) in AWMN, web proxies, FTP or shared disk servers, XMPP instant messaging servers, IRC servers, cloud services as the \textit{Cloudy} distribution~\cite{Selimi:2015:CSG:2852375.2852752} in guifi.net, Internet cube, BitTorrent tracker, IndeCP or Internet service in FFDN, private VoIP service and weather monitoring in Sarantaporo.gr.

\vspace{2mm}
\subsection{Lawful framework of operation}
\label{s:lawful}

An operational framework of CNs (legal status, rights, obligations) which is not well defined may impede the  attraction of new participants. The level of support of CN initiatives by the state or local administration has an impact on users' decisions to join or not the network~\cite{abdelaal2013social}. When local authorities or another third-party organization with clear legal status are involved, \eg by signing licenses, the user's concerns are easier overcome and the decision to participate looks far less risky. The response of most CN initiatives to these reservations is to develop legal entities, and set forth licenses and agreements as legal documents specifying the terms and conditions of participation in the network. 
  
\subsubsection{Operation as legal entities}
The majority of CNs have developed legal entities to represent the network to third parties (Table \ref{tab:cn_aspects}). For example, Guifi.net created the guifi.net Foundation, AWMN the Association of AWMN, FFDN consists of non-profit member organizations registered as telecom operators, Sarantaporo.gr operates as a non-profit civil partnership subject to the Greek legal framework about NPOs, Freifunk has the Forderverein freie Netzwerke e.V. as a reference NPO authority, TakNet is a social enterprise and B4RN a community benefit society.

\subsubsection{Licenses and Agreements}
\label{subsec:licences}

Besides the legal status, CNs normally make use of legal documents, such as Licenses and Agreements, to specify the frame of their members' participation and their own interaction with third-party entities. 

Guifi.net and FFDN utilize a Network Commons License (NCL) for establishing the rights and duties of subscribed participants. Moreover, guifi.net has developed collaboration agreements (\textit{Type A}, \textit{Type B}, \textit{Type C}) that define the terms of conditions of third party collaboration within the network. Any private sector entity that wants to perform economic activities and use a significant amount of resources of the network has to sign an Agreement with the Foundation and participate in the compensation system (\cref{subsec:compensation}). Freifunk uses the PicoPeering Agreement that promotes the free exchange of data within the network 
Ninux participants comply with the Ninux manifesto, which is a variation of the PicoPeering Agreement.

\vspace{2mm}
\subsection{Incentive mechanism classification}
Several of the incentive mechanisms that are described in sections from \ref{s:enforcing} to \ref{s:lawful}  have never gone beyond the paper analysis stage. On the other hand, several others are indeed applied in existing CNs. The financial compensation system of guifi.net, the social events, meetings and workshops organized by many CNs, the adoption of licences in Freifunk and guifi.net, as well as the introduction of a lawful operational framework serve, one way or another, as incentive mechanisms that motivate the participation of different types of stakeholders in CN initiatives, as shown in Table \ref{tab:mechanism_stakeholder}. 

Some of these incentive mechanisms apply almost invariably to all CNs. The lawful operational status, for example, is mandatory if the CN wants to attract critical masses of users, but also private sector entities and the support from public agencies. Equally common among CNs is the care for social events and meetings that can strengthen the links between their members and satisfy socio-cultural motives of users. On the contrary, incentive mechanisms of economical nature, such as the financial compensation scheme and the donation certificates issued bu guifi.net for tax deduction purposes are more relevant in CNs that support commercial operations over them.

For sure, it would be rather wise to match the incentive mechanisms with the different stakeholder types. Hence, volunteers would be more responsive to incentive mechanisms that underline political and cultural causes; private sector service providers would respond, maybe exclusively, to incentive mechanisms with economic implications; and local authorities will be much more prone to get involved when they realize that public expenses can be saved or some political strategic objective be served through this involvement.

By far, the majority of incentive mechanisms target CN users. One aspect that is not well understood is how the effectiveness of a mechanism varies with different features of the community; namely, if we could have a characterization of a community according to a fixed set of attributes (urban vs. rural, educational level, professional background, dominant political preferences) that could predict which incentive mechanism would best mobilize its members. An important parameter in this context is the size of the community. Characterizations along attributes is easier if the community is small\footnote{But not too small. The CN will not be sustainable if there are not enough human resources to pull from.} and with roughly uniform interests and professional background. As their size grows, such characterizations become harder and so does any attempt to predict the suitability of incentive mechanisms.

\section{Discussion and conclusions} \label{closure}

The survey has examined the issue of sustainability in community networks and the various aspects within political, socio-cultural and economic perspectives for CN participants. Special focus has been given to the economic sustainability of CNs. Open access network infrastructure models were presented and compared to tradition telecom business models while possible funding options from private and public entities and CN members were analyzed. Economic sustainability is a challenging issue which seems to be a necessary but not the only sufficient condition for approaching a sustainable operation of the CN. Socio-cultural and politic perspectives are needed as well. 

Chosen CNs were studied with respect to actual and theoretical mechanisms for enhancing sustainability by organizing and encouraging member participation. Each CN is composed by four basic types of entities (\ie volunteers, private sector entities, users, public agencies). These entities experience their own motives for joining the network and take part in mechanisms deployed to organize their contributions. In order to match their political, socio-cultural and economic interests, corresponding mechanisms have to be in place.

It appears that sustainability cannot be reached following a set of exhaustive rules and there are no clearcut answers for approaching it. However, checkpoints or indicative guidelines can be used to assess it. These can be summarized in the following evaluation form following the fieldwork in~\cite{NETCOMMONS_D2_2}. 
\subsection{Economy}
\subsubsection{Market and model of provision}
\begin{itemize}
\item To which extent is the community network supported by non-profit/community based network access and services provision?

\item To which extent does the community network rely on a commercial provider? What is the nature of this provider (e.g. for-profit vs. social enterprise, or local vs. non-local)?

\item To which extent does the model of network provision of the community network face competition from commercial for-profit telcos on the basis of quality of signal/provision, lower cost and/or better network maintenance?
\end{itemize}

\subsubsection{Resources}
\begin{itemize}
\item To which extent does the community network manage to survive economically, i.e. to afford the necessary hardware and labour-power necessary for running the network?
\item To which extent can the community network ensure that it has enough resources, supporters, workers, volunteers, and users?
\item To which extent does the community network rely on internal funding sources?
\item To which extent does the community network rely on external funding sources?
How regular are they?
\item Are there possibilities for the community network to obtain public or municipal
funding or to co-operate with municipalities, public institutions or the state in providing access and services? 



\item To which extent does the community network rely on a single individual or a small group of actors for providing the necessary resources (time, skills, money)?


\end{itemize}



\subsubsection{Network wealth for all}
\begin{itemize}
\item To which extent does the community network provide gratis/cheap/affordable network and Internet access for all?

\item If subscriptions are used, are they affordable?
\item To which extent are there different pricing schemes such as for residential users,
small enterprises, bigger firms, and public institutions (e.g. schools)?
\item How can the community avoid or lower the digital divide? 

\item What technological skills are required of the average user to benefit from the community network?
\end{itemize}

\subsubsection{Needs}
\begin{itemize}
\item To which extent are the community needs served by the community network? 
\item To which extent are the needs of diverse individuals (e.g. by gender, age, nationality) and groups in the community served by the community network?
\item To which extent are the needs of local businesses served by the community
network?
\end{itemize}

\subsection{Politics}
\subsubsection{Participation/governance}
\begin{itemize}
\item How is the community network governed? How does it decide on which rules, standards, licences, etc. are adopted? 
\item To what extent does the community network allow and encourage the participation of community members in governance processes?
\item To what extent are there in place mechanisms for conflict resolution and for proceedings in the case of the violation of community rules?

\end{itemize}

\subsubsection{Data ownership and control}

\begin{itemize}
\item To which extent does the community network enhance the protection of privacy of user data?

\item To which extent does the community network provide opportunities for active user
involvement in the management of their data? What are the skills required and how are they provided?

\item To which extent and for how long are user data kept in servers controlled centrally
(e.g. by the network administrators)? How do you guarantee that data storage is
done in line with data protection regulation and is privacy-friendly?

\end{itemize}

\subsection{Culture}

\subsubsection{Community spirit} 

\begin{itemize}

\item How closely knit is the community? To which extent are trust and solidarity present and how are they manifested?

\item To which degree is the community network a “geek public” that has an elitist, exclusionary culture or a “community public” that is based on a culture of unity in diversity? 

\item To which extent does the community network provide mechanisms for learning,
education, training, communication, conversations, community engagement, strong democracy, participation, co-operation, and well-being? In what ways?

\item To which degree is the community network able to foster a culture of togetherness and conviviality that brings together people? In what ways? 
\end{itemize}

\subsection{Future research directions}

Having looked carefully into the broader issue of sustainability, our future work is going to focus in specific CNs and will seek to propose incentive mechanisms that can address the sustainability challenge. This is going to be pursued through different directions. 

One direction is on the effort to import elements from the guifi.net model such as the involvement of professionals in the network and the provision of commercial services over it. More specifically, part of the work will be devoted to analyzing the incentive mechanisms that guifi.net has put in place: the sustainability of its compensation system, which is the main tool for incentivizing the participation of commercial entities in the CN; and the exportability of this model to the newer and promising Sarantaporo.gr CN. 

A second, related direction, is through the launch of an open source mobile application over the CN that realizes mobile crowdsourcing and sharing economy practices. We will analyze incentives (\ie game theoretic tools, reciprocity theories) that can be embedded in the application to maximize its use, and through that, the use of the CN. Following the prototype of the application released in~\cite{NETCOMMONS_D3_2}, we will then design incentives to accommodate the agricultural services that are of interest in this case, taking into account the particularities of the CN network such as the ownership of network nodes and network connectivity alternatives that are available in the area.

\section{ACKNOWLEDGMENTS}
The authors acknowledge the support of the European Commission through the Horizon 2020 project netCommons (Contract number 688768, duration 2016-2018).






\bibliographystyle{plain}
\bibliography{bibtex/bib/references,bibtex/bib/deliverables}




%




\end{document}